\documentclass{aa}

\bibpunct{(}{)}{;}{a}{}{,} 

\usepackage{graphicx}
\usepackage[varg]{txfonts}
\usepackage{amsmath}

\usepackage{multirow}

\newcommand{\etal}{{\em et al.}}

\newcommand{\hms}[3]{$#1^\mathrm{h}#2^\mathrm{m}#3^\mathrm{s}$}
\newcommand{\hm}[2]{$#1^\mathrm{h}#2^\mathrm{m}$}

\begin{document}
\title{Analysis of on-sky MOAO performance of CANARY using natural guide stars}

\author{Fabrice Vidal \inst{1} 
\and Eric Gendron  \inst{1} 
\and G\'erard Rousset \inst{1} 
\and Tim Morris \inst{2}
\and Alastair Basden \inst{2}
\and Richard Myers \inst{2}
\and Matthieu Brangier \inst{1}
\and Fanny Chemla \inst{3}
\and Nigel Dipper \inst{2}
\and Damien Gratadour \inst{1}
\and David Henry \inst{4}
\and Zoltan Hubert \inst{1}
\and Andy Longmore \inst{4}
\and Olivier Martin \inst{1}
\and Gordon Talbot \inst{2} 
\and Eddy Younger \inst{2}
}

\institute{
LESIA, Observatoire de Paris -- CNRS -- UPMC -- Universit\'e Paris Diderot, 5, Place Jules Janssen, 92190 Meudon, France   \\
\email{fabrice.vidal@obspm.fr}
\and Centre for Advanced Instrumentation, Durham University, South Road, Durham DH1 3LE, UK
\and GEPI, Observatoire de Paris -- CNRS -- Universit\'e Paris Diderot, 5, Place Jules Janssen, 92190 Meudon, France   \\
\and UKATC, Royal Observatory Edinburgh, Blackford Hill, Edinburgh EH9 3HJ, UK 
}

\abstract
{The first on-sky results obtained by CANARY, the Multi-Object Adaptive Optics (MOAO) demonstrator, are analysed. The data were recorded at the William Herschel Telescope, at the end of September 2010. We describe the command and calibrations algorithms used during the run and present the observing conditions. The processed data are MOAO-loop engaged or disengaged slope buffers, comprising the synchronised measurements of the four Natural Guide Stars (NGS) wavefront sensors running in parallel, and near Infra-Red (IR) images. We describe the method we use to establish the error budget of CANARY. We are able to evaluate the tomographic and the open loop errors, having median values around 216~nm and 110~nm respectively. In addition, we identify an unexpected residual quasi-static field aberration term of mean value 110~nm. We present the detailed error budget analysed for three sets of data for  three different asterisms.  We compare the experimental budgets with the numerically simulated ones and demonstrate a good agreement. We find also a good agreement between the computed error budget from the slope buffers and the measured Strehl ratio on the IR images, ranging between 10\% and 20\% at 1\,530~nm. These results make us confident in our ability to establish the error budget of future MOAO instruments. }

\maketitle



\section{Introduction}
The understanding of the physics and of the formation of high redshift galaxies requires the multiplexed observations of very large number of targets with spatial and spectral resolution capabilities. Because of their faintness this science case also requires the photon-collecting area of an extremely large telescope (ELT) of 30-40~m class. 
During the E-ELT instrument phase A studies several MOS concepts were studied, amongst which EAGLE \citep{Cuby10}, a near infrared multi-integral field spectrograph fed by Multi-Object Adaptive Optics (MOAO) \citep{Hammer02}. 

EAGLE is a near-infrared (IR) multi-object integral field spectrograph with 37.5~mas spatial sampling and a spectral resolution of 4\,000. It aims to simultaneously analyse 20 targets in a wide field of view (FOV) of about 7.5 arcmin in diameter. Thanks to the small angular size of the high-z galaxies,  adaptive optics (AO) is required \citep{Puech2010}.
Indeed, all the ground-based large telescopes suffer from spatial resolution degradations due to atmospheric turbulence, leading to an image full width at half maximum (FWHM) of the order of 1 arcsec. To overcome these effects, AO is required to compensate for the wavefront distortions in real time as demonstrated at the end of the 1980s in astronomy \citep{Rousset90}. However, conventional AO is only able to compensate for turbulence across a relatively small FOV of the order of a few tens of arc seconds in the near IR (H band for instance) due to the anisoplanatism effect. Therefore new AO technologies are required to overcome this strong limitation. Multiconjugate AO (MCAO) \citep{beckersMcao1988,Marchetti08} and Ground Layer AO (GLAO) 
\citep{RigautVeniceGlao02,Milton08} are novel AO system concepts that offer larger compensated FOV. However MCAO can deliver a good correction only in a FOV of the order of 1 or 2 arcmin while GLAO delivers a moderate correction, with a typical gain of a factor of 2 in FWHM with respect to the seeing \citep{bendekMMT2011, Tokovinin2012}, but in a much wider FOV up to the order of 10 arcmin.

It is not possible to design and implement an MCAO system covering such a wide FOV because of the conservation of beam \'etendue. Using the specific case of a high redshift galaxy programme, only the galaxies themselves are of interest for the AO correction, not the continuous FOV. In addition, their angular extent is very limited to the order of 1 or 2 arcsec. The multi-object AO (MOAO) concept was initially proposed to tackle this problem \citep{Hammer02}. It aims to simultaneously compensate the turbulence for a large number of very faint small science objects distributed over a wide FOV. For that purpose one deformable mirror (DM) is implemented per target in a dedicated optical train feeding each integral field unit of a spectrograph. Moreover, the galaxies are  too faint to be able to measure any wavefront distortion in real time. It is therefore necessary to find a number of bright guide stars (GS) within the wide instrumental FOV for that purpose. These guide stars can be natural GS (NGS), but for questions of sky coverage, laser GS (LGS) are needed. The light from each GS must be picked-off in the FOV and sent to a wavefront sensor (WFS).  In such a configuration, the WFSs do not see any feedback from the correction applied on the DMs; they are thus working in an open loop mode. Finally all the WFS measurements made across the FOV must be mutually processed in order to reconstruct the 3D turbulent volume above the telescope. 
This is achieved using a tomographic approach \citep{Ragazzoni99}. Then a projection of the volume in the direction of each target \citep{Fusco01,Vidal10JOSA} allows us to compute the correction to be applied to each DM in the system in open loop. 
A possible configuration for such a MOAO system is the one envisioned for the EAGLE instrument of the E-ELT \citep{Rousset10}. It makes use of 6 LGSs at the periphery of the FOV and between 4 to 6 additional NGSs selected in the field. The two main critical features of MOAO are tomographic reconstruction of the turbulence and the DM operation in open loop. 

In order to improve the technology readiness of the EAGLE instrument, a pathfinder was proposed to demonstrate the feasibility and the performance of MOAO \citep{Myers08}: this is the CANARY project. A first on-sky testing of on-axis open-loop command has been already reported \citep{Volt08} as well as first MOAO laboratory demonstrations \citep{Ammons10, Vidal10AO4ELT}. The next step is to demonstrate on-sky the capability of the open loop tomography using a number of in-the-field GS, both laser and natural. This is the main goal of CANARY.

Tomography has been successfully demonstrated in MCAO with MAD at VLT on NGS only \citep{Marchetti08} and more recently with a 5 LGS constellation on GeMS at Gemini Observatory \citep{Rigaut2013_Gems_1, Neichel2014_Gems_2}.
However, the MCAO approach uses multiple DMs conjugated at different altitude in the optical train of the instrument. This leads to a specificity of MCAO: the tomography problem is partially optically solved. This maybe directly taken into account by regularizing the pseudo-inverse of the interaction matrix between the WFSs and the DMs. In MCAO, Minimum Mean Square Error (MMSE) reconstructors without the turbulence profile knowledge still allows us to achieve a good level of performance \citep{Vidal2013}. In MOAO, with one DM per target the least-squares approach only leads to a partial correction of the wavefront, i.e. correcting only the ground layer. In fact, the knowledge of the turbulence profile is not crucial in MCAO but becomes a limiting factor in MOAO and is even more dramatic as the GS can be positioned farther off-axis from the target object.

Considering the open loop operation, typical DM errors such as creep and hysteresis have been studied earlier \citep{Morzinski2008, kellererDM2012}. For certain DM types including that in CANARY, they have been demonstrated to be small compared to the fitting error. In addition specific open loop procedures have to be developed as presented in section \ref{ControlAndCalibration} for CANARY.

In Sect.~\ref{instrument}, we present the CANARY instrument with a hardware and software description and in section ~\ref{ControlAndCalibration} the control and calibration algorithms. The data reduction approach is described in Sect.~\ref{reduc}. In particular, the error budget computation is detailed. Section~\ref{OnskyResults} presents the on-sky results, the observing conditions and introduce the three selected NGS asterisms. The data are processed to retrieve a 15 layer $C_n^2(h)$ profiles in order to quantify the tomographic and open loop errors. Section~\ref{details} gives the detailed error budgets established for three sets of data during one night of observations. The results are compared to numerical simulations. We conclude in Sect.~\ref{conclusion}.

\section{The CANARY instrument} 
\label{instrument}
\subsection{Introduction}

\begin{figure}
\centering
\includegraphics[width=8cm]{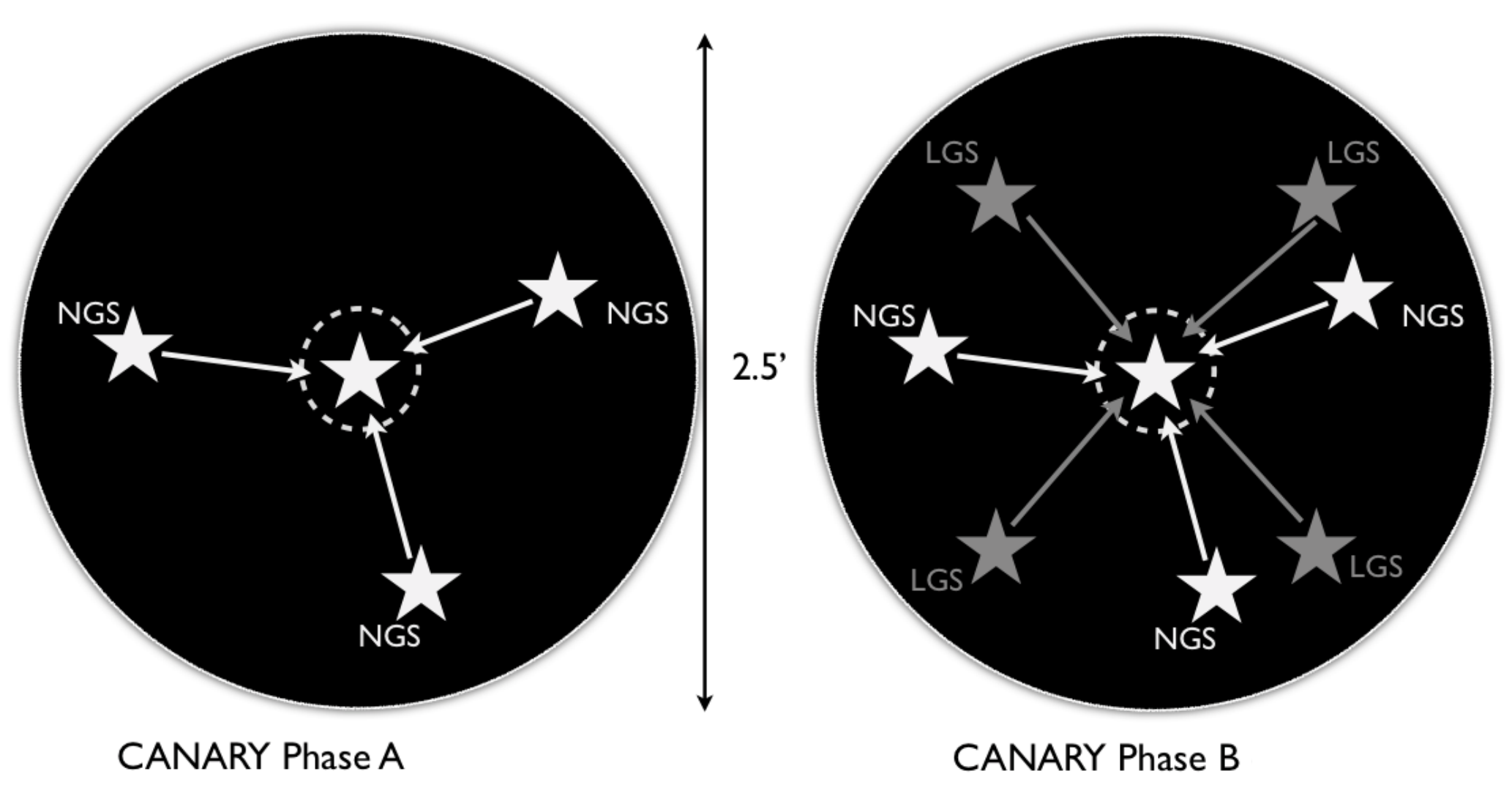}
\caption{
CANARY MOAO demonstration configuration. Phase A of the system (left) uses three off-axis NGS to reconstruct the wavefront while one deformable mirror is driven in open loop on the on-axis star. Phase B (right) added 4 Rayleigh LGS in addition of the three NGS. 
}
\label{CANARY}
\end{figure}

The CANARY pathfinder 
implements a single target channel of a MOAO system and is deployed at one of the Nasmyth foci of the 4.2m William Herschel Telescope (WHT) at the Roque de Los Muchachos Observatory on the island of La Palma in Spain. CANARY is a fast track experiment. The design phase started beginning of 2008 while the first light happened in September 2010 \citep{Gendron11}. 
CANARY is a project planned in three phases (respectively called A, B, C), each phase of increasing system complexity \citep{Myers08}. 
It will lead to a comprehensive demonstration of the MOAO configuration as foreseen for EAGLE on the E-ELT  \citep{Rousset10}.
Phase A makes first use of three open-loop off-axis NGS WFSs and one on-axis open-loop-controlled DM of $8\times8$ actuator array. Phase B will use, in addition, four Rayleigh LGS to augment the turbulence tomography in open loop (see Fig.~\ref{CANARY}). Phase C will introduce the full configuration proposed for EAGLE. We will use a  woofer DM in closed-loop (in fact the 8x8 actuator DM), as a first stage simulating M4 of the E-ELT \citep{VernetM4_2012} with an open-loop tweeter DM as a second stage simulating the MOAO channel. The tweeter DM in Phase C will be a higher-order DM from ALPAO with $17\times17$ actuators (241 useful). In order to reconfigure the system between the three phases, the optical design is build around a set of interchangeable optical modules. The Rayleigh LGSs used for these demonstrations have variable range-gate height and extension in order to simulate many of the LGS effects that will be encountered at the E-ELT.

\subsection{Phase A system}

\begin{figure}
\centering
\includegraphics[width=9cm]{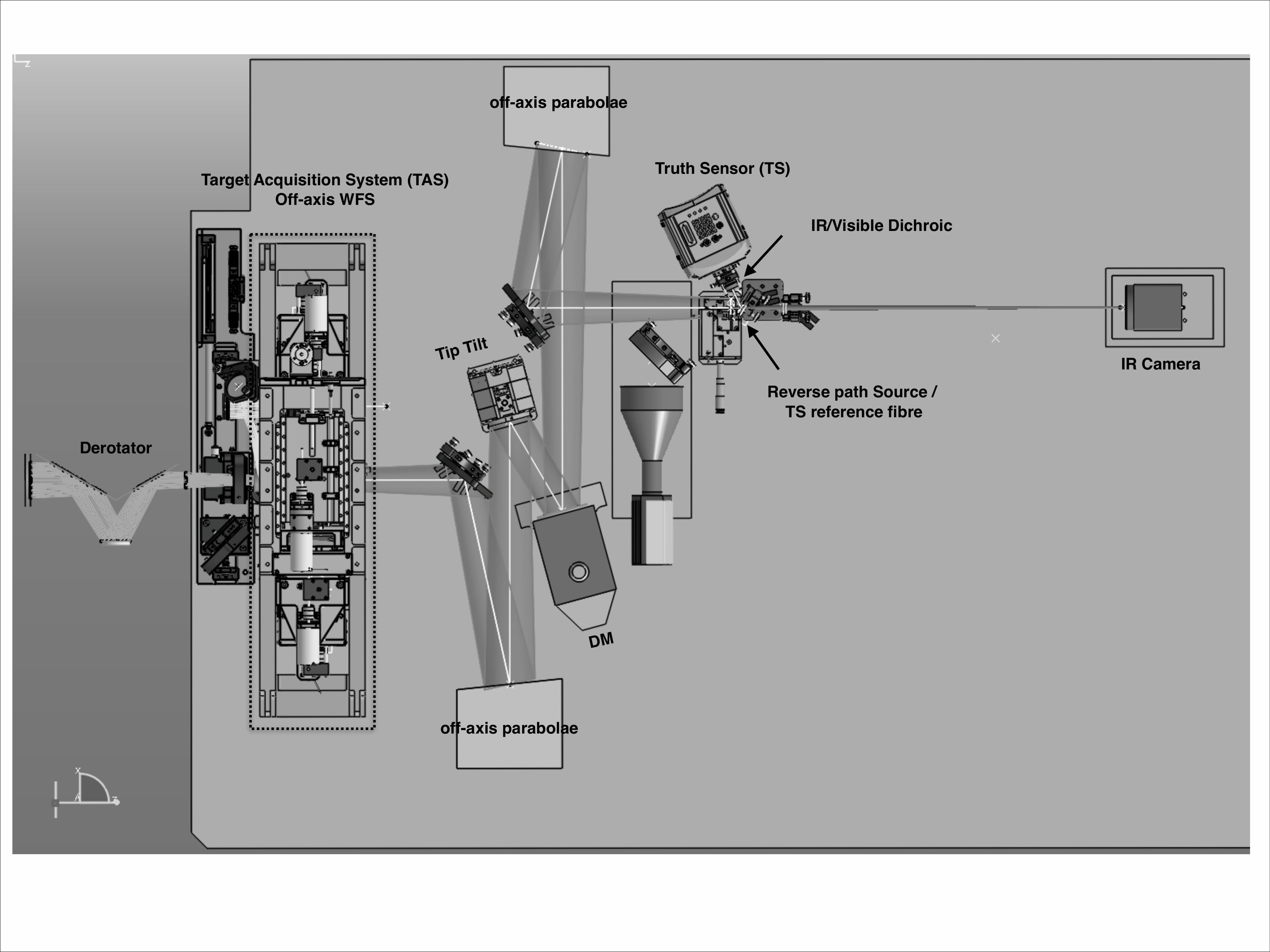}
\caption{The CANARY input focal plane is equipped with a Target Acquisition System (TAS) carrying 3 off-axis WFS working in open loop. Light from the central star is transported by the off-axis parabolic relay including the Tip-Tilt stage (TT) and the deformable mirror (DM) to the output corrected focus. At this focus, the light is reflected in the visible by a dichroic plate to a fourth WFS called the Truth Sensor. Finally, the IR light is transmitted to the IR camera observing at a central wavelength of $\lambda=1\,530$~nm.}
\label{CANARYBench}
\end{figure}

\begin{figure}
\centering
\includegraphics[width=9cm]{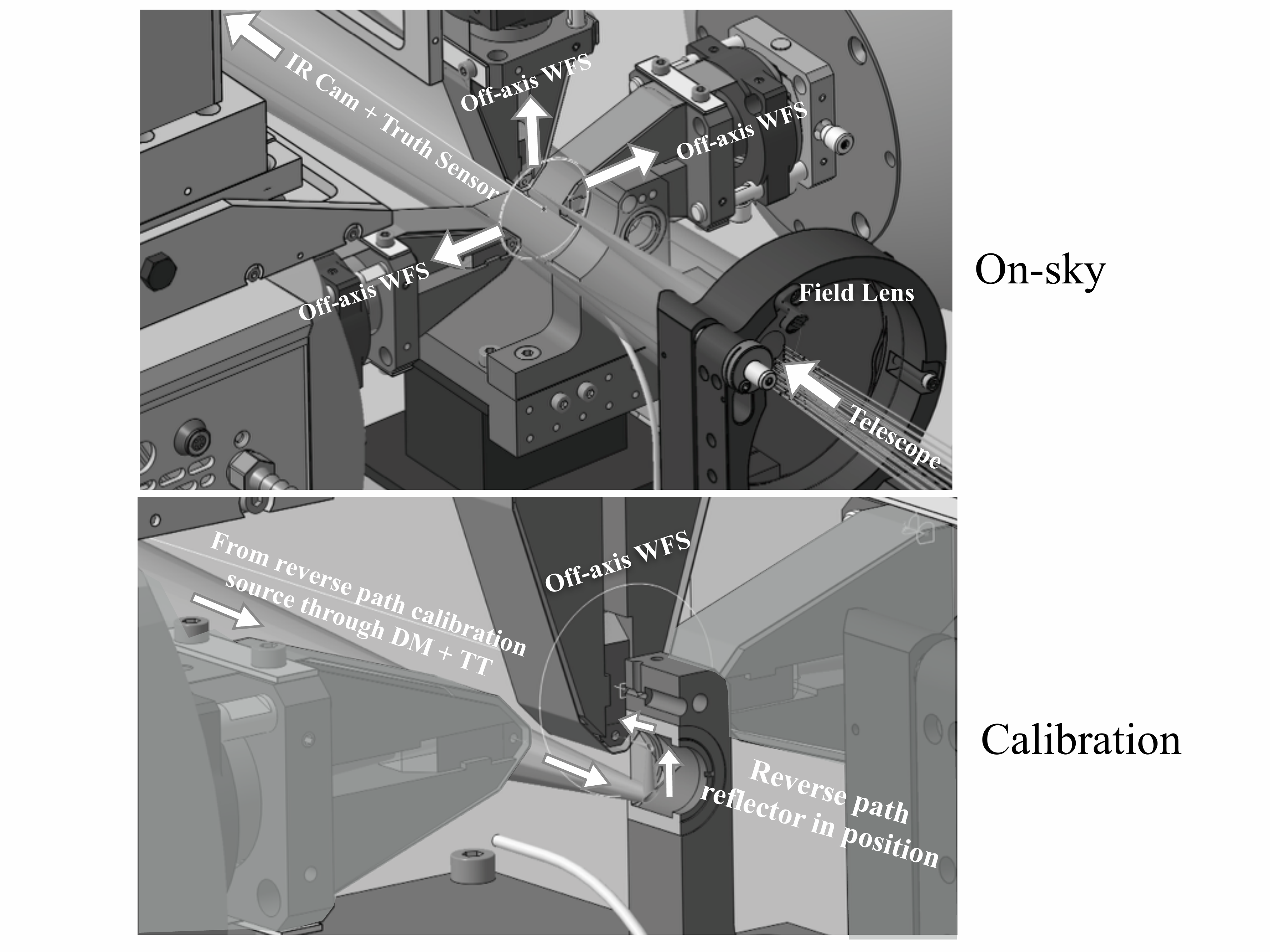}
\caption{Zoom of the CANARY focal plane. 
Upper part illustrates the on-sky configuration where the off-axis light is reflected to the off-axis WFS. The light from the central star is sent via the DM and TT to the IR camera and TS. 
Lower part illustrates the open-loop calibration configuration using the reverse path calibration source. The light is illuminating the DM and TT and is retroreflected at the focal plane on one off-axis WFS. This allows us to retrieve interaction matrices for the off-axis WFS and compute misalignment and scaling errors between all the WFS.}
\label{FocalPlaneZoom}
\end{figure}

At the entrance of the bench, upstream of the Nasmyth focal plane, the field rotation is compensated using a K-mirror derotator.
CANARY is equipped with three WFS that are able to patrol the derotated field ($2.5\arcmin$ in diameter, plate scale  0.22~mm~arcsec$^{-1}$) and acquire the selected off-axis NGSs. The star that mimics the corrected science target is situated at the centre of the field. 
This on-axis central star is observed by an IR imaging camera and a fourth WFS called \emph{Truth Sensor} (TS) which measures the DM-corrected wavefront (Figs ~\ref{CANARYBench} and ~\ref{FocalPlaneZoom}).

The four WFS are identical, of Shack-Hartmann (SH) type, with $7\times7$ sub-apertures (only 36 illuminated). 
The total field of view per WFS sub-aperture is close to 4\arcsec.
They use Andor iXonEM 860 EMCCD cameras with $24~\mu$m pixel size, 
featuring $128\times128$ pixels. The measured read-out noise is 0.3, 0.5, 0.55 and 0.7~e$^-$~rms per pixel for the four cameras due to the Electron Multiplication gain. We use $16\times16$ pixels of $\approx 0.26\arcsec$ scale in each SH sub-aperture. 
The minimum distance between two off-axis WFSs is $20\arcsec$, limited by both hardware and software anti-collision systems. 

The central star beam is sent to the performance diagnostic IR camera and TS via a two off-axis parabolic mirror relay. 
The relay includes the CILAS $8\times8$ piezostack array DM conjugated to the WHT pupil. This DM was manufactured in 1991 and previously used at ESO in the ADONIS instrument \citep{beuzit94}.  It is made in a very hard piezoelectric material and exhibits hysteresis and creep effects \citep{kellererDM2012}.
Wavefront errors caused by ``creeping'' of the DM surface are observed but their effet on performance is minimal.
The high-speed tip-tilt mirror is a copy of that used in the VLT instrument SPHERE \citep{Sphere2008}. The mirror pointing is controlled by a servo-loop based on internal position sensors  to minimise open-loop error.

At the output of the relay, the beam is split by a dichroic plate: the transmitted IR flux (1 to 2.5~$\mu$m) is sent to the IR camera. This is a Xeva-1.7-320 from Xenics, featuring $320\times256$ pixels of $30~\mu$m pitch, 
with a measured readout noise of $200\pm30~\mathrm{e}^-$ rms per pixel and a gain of $10~\mathrm{e}^-  \ \mathrm{ADU}^{-1}$. 
The plate scale of the IR camera is $0.0371\arcsec$ per pixel. The combination of the H filter bandwidth, the QE efficiency of the camera and the atmosphere absorption gives an effective wavelength of the IR image centered at $\lambda=1\,530$~nm with a bandwidth of $\Delta \lambda = 160$~nm.
The visible part of the spectrum (below 900 nm) is reflected by the dichroic and goes to the TS. The TS allows us to check the efficiency of the turbulence compensation in real-time, to perform system calibration tasks and to close the loop for performance comparison between conventional closed loop AO (hereafter called Single Conjugate AO: SCAO) and open loop MOAO.

A telescope, turbulence and star simulator allow us to fully characterise the system when under testing in the laboratory or during daytime maintenance at the telescope. A deployable acquisition camera can be inserted to observe the whole field (2.5~arcmin diameter) for calibration of the telescope pointing and guide star acquisition.
The entrance focal plane of CANARY is also equipped with deployable calibration sources of various diameters to emulate both seeing or diffraction-limited sources. In addition, a specific feature is the so-called reverse-path calibration source, discussed in Sect.~\ref{interactionmatrix}. A more detailed overview of the CANARY can be found in \citep{Gendron10}.

\subsection{Real-time computer}
\label{wpurtc}

The real-time computer (RTC), called DARC (Durham Adaptive optics Real time Controller) is described in
\citep{Dipper10} and \citep{Basden2010}. 
The system is driven at a selectable sampling frequency of up to 250~Hz, limited by the WFS camera readout rate.
The four WFS cameras are synchronised and read through dedicated FPGA-based hardware. The AO controller is CPU based, and optimised for multi-threaded operation on multi-CPU architectures. It exhibits $0.7$ ms latency between the latest readout pixels of the WFSs and the time when the DM actuators are reaching half-stroke value. DARC also allows clients to obtain continuous or sub-sampled real-time data streams of SH images, slopes, and DM command voltages. DARC also features a number of different real-time algorithms. 
The SH image processing algorithm uses adaptive windowing ($12\times12 $ pixels) of the SH spots on a sub-aperture-by-sub-aperture basis, together with centroiding done on 10 brightest pixels \citep{basdenBrightestPix2012}. These features allowed us to cope with the dynamics required for open-loop wavefront sensing depending on the observing conditions and to reduce the effect of readout noise. 

\subsection{Loop delay}
We measured the loop delay by introducing a known, white noise, voltage pattern on the DM command. By recording synchronised sets of TS slopes and DM voltages, we can deduce the delay from the WFS measurements. We measured a 1.5 and 1.9 frame delay at 150 and 250~Hz respectively. We define  delay as the time taken from the beginning of the WFS integration to the application of the corresponding command on the DM. 

\section{Control and calibrations algorithms}
\label{ControlAndCalibration}
\subsection{Interaction matrices and deviations}
\label{interactionmatrix}
The interaction matrix between the DM and the truth sensor is an important calibration item in the control of CANARY.
As in conventional AO, the interaction matrices reflect the optical relationship between the DM and the WFSs \citep{Boyer90}. They are  measured experimentally by actuating the mirror and measuring the impact on a given wave-front sensor. 
They are even more important in our open-loop scheme, since we are now concerned with not only the relative amplitude between coefficients, but also with their {\em absolute} amplitude. Whereas a global multiplicative factor would just act as a loop gain when operating in closed loop, here it will operate as a scaling factor on the correction.
In CANARY, we have chosen to simultaneously activate all actuators \citep{kellererDM2012,Marchetti08} with a sinusoidal wave pattern using a specific temporal frequency for each and retrieve the matrix from the WFS measurements using a lock-in approach. This particular method has been demonstrated \citep{Vidal2009, kellererDM2012} to be one of the best for identifying the DM model. The method has also demonstrated excellent on-sky capabilities for interaction matrix identification with the TS, the quality of the non-zero matrix coefficients only differs by 2\% from the laboratory ones.
We will call $M_i$ the interaction matrix between DM and TS, and $M_c$ its generalised inverse. The matrix $M_c$ is computed using a singular value decomposition of $M_i$. We observe a significant drop-off of the last eigenvalues and filtered the last 7 modes to keep a conditioning number of 50.

In an open-loop scheme, the focal plane off-axis wave-front sensors have no optical feedback from the deformable mirror. This, in turns, means that no interaction matrix can be measured physically, since the WFS do not see the DM. This limitation, inherent to the MOAO scheme, is a severe drawback since the optical relation between wave-front sensors and DM does exist within the tomographic reconstructor. That is why CANARY is equipped with a reverse path calibration source, which is a seeing-limited source illuminating the DM from the output focal plane to one of the open-loop WFS located in the input focal plane.
This is achieved using a retroreflector system, which preserves the pupil orientation (\ref{FocalPlaneZoom}). Therefore we are able to record interaction matrices with each off-axis WFS, despite the open-loop scheme. Those matrices are the starting point for finding out all instrumental model parameters which are: pupil image translations and magnifications, rotations of the lenslet arrays and of the WFS cameras, and WFS sensitivities. This tool was first used for the purpose of fine alignment and also for final calibration of all the WFS parameters \citep{Vidal10JOSA, Brangier2012}. These parameters could then be taken into account in the computation of the tomographic reconstructor. 

\subsection{The Learn \& Apply tomography algorithm}
\label{LandA}

The linear tomographic reconstruction matrix is derived using the \emph{Learn \& Apply} (L\&A)  algorithm from \citep{Vidal10JOSA}, where the minimum mean square error (MMSE) reconstructor is computed from an atmospheric model directly identified on-sky from the WFS measurements.
Engaging the open-loop compensation is preceded by the acquisition of an on-sky WFS data set, from which the optimized estimator is deduced.
We detail hereafter the implementation.

We define notations as follows. 
The set of local wavefront slopes along $x$ and $y$ directions, measured by a WFS $i$ with $p$ sub-apertures, is collected into a slope vector $\vec{S_i}$ 
$$ \vec{S_i} = \left(
     \begin{array}{cc}
	sx_1 & \\
	\vdots &\\
	sx_p &\\	
	sy_1 & \\	
	\vdots &\\
	sy_p &\\	
     \end{array} 
     \right) - \vec{S}_{ref\ i}
$$
After acquisition of number of frames \textit{nframe} of an on-sky WFS data set we build a matrix $M_{offaxis}$ by concatenating horizontally the column-vectors formed by synchronous measurements $t=[1, ... , nframe]$ from the $n$ off-axis WFSs. Each WFS slope is subtracted from its own reference slope ($\vec{S}_{ref\ i}$) including all the static aberrations for the observing direction of the NGS $i$ (see Section \ref{RefslopesCal}). 
$$
     {M_{offaxis}} = \left(
     \begin{array}{ccc}
	\vec{S}_{1\ t=1} & \cdots & \vec{S}_{1\ t=nframe}\\
	\vec{S}_{2\ t=1} & \cdots & \vec{S}_{2\ t=nframe}\\
	\vdots & \vdots &\vdots\\
	\vec{S}_{n\ t=1} & \cdots & \vec{S}_{n\ t=nframe}
     \end{array}
     \right).
 $$
Similarly, in the same time, synchronous slope vector measurements from the on-axis truth sensor are concatenated in the same way to form the matrix $M_{{central}}$ 
$$
     {M_{central}} = \left(
     \begin{array}{cccccccc}
	\vec{S}_{central\ t=1} & \cdots & \vec{S}_{central\ t=nframe}\\
     \end{array}
     \right).
$$
\citep{Vidal10JOSA} infers that for this particular turbulence sequence the best tomographic estimator $M_t$ is the one that directly links inputs and outputs (i.e. $M_{offaxis}$ and $M_{central}$) as
\begin{equation}
M_{central} = M_t.M_{offaxis} \ ,
\label{equationTomo}
\end{equation}
and solving, in a least-squares sense, this equation for $M_t$ is given by \citep{Vidal10JOSA}
\begin{equation}
M_t =  (M_{central} {M^t_{offaxis}}) (M_{offaxis}{M^t_{offaxis}})^{-1}\ .
\label{equationTomoSol}
\end{equation}

It is noticeable that this estimator $M_t$ tends towards the MMSE estimator as the acquisition time tends towards infinity (and under hypothesis of stationarity), because both matrices $(M_{offaxis} {M^t_{offaxis}})$ and $(M_{central}{M^t_{offaxis}})$ tend towards the covariance matrix of the sensor measurements. 
We respectively call these matrices $C_{OffOff}$ and $C_{OnOff}$.
Then we can write Eq.~\ref{equationTomoSol} as
\begin{equation}
M_t =  C_{OnOff} \, . \, C^{-1}_{OffOff} \ .
\label{RetrieveTomo}
\end{equation}

The covariance between two elementary slopes in $x$ of sub-apertures $\theta \in  [1, ..., p] $ and $\nu~\in~[1,...,m]$  of respectively two WFS $i$ and $j$ is noted $<s_{x\theta} \, s_{x\nu}>_{ij}$. Here $p$ and $m$ are respectively the total number of subapertures for WFS $i$ and $j$. The covariance value can be theoretically computed from the Kolomogorov covariance maps~: expressions in direct space have been given by \citep{butterleySLODAR} and some analytical approximations have recently been given by \citep{martinSpie2012}.
We expressed covariance
in the Fourier domain for a single turbulent layer as
\begin{multline} 
\mathcal{F} ( <s_{x\theta} \vec(k) s_{x\nu} \vec(k)>_{ij})   \propto   k_x^2 
\,  \tilde{\Pi}_i(\vec{k})   \,  \tilde{\Pi}^*_j(\vec{k})  \,
r_0(h_l)^{-\frac{5}{3}} \,  \| \vec{k} \|^{-\frac{11}{3}} \\
\times
e^{ -2i\pi \left[   k_x (h_l(\alpha_i - \alpha_j)  + x_{i\theta} - x_{j\nu}  ) +    k_y (h_l(\beta_i - \beta_j)  + y_{i\theta} - y_{j\nu}  ) \right]    } 
\label{ExpressionCovariancegen}
\end{multline}
with $r_0 (h_l)$ the Fried parameter at altitude $h_l$,  $\vec{k} = (k_x,k_y)$ the conjugate Fourier variable of the separation $\vec{r} = (x,y)$  between  two  sub-apertures $i$ and $j$, and $\tilde{\Pi}_i(\vec{k})$ the Fourier transform of the sub-aperture shape function $\Pi(\vec{r})$ (i.e. equal to 1 within the sub-aperture and 0 elsewhere). The separation between two WFS is characterised by the difference in their pointing directions ($\alpha_i,\beta_i$) and ($\alpha_j,\beta_j$), and the altitude $h_l$ of the considered layer. 

During a real scientific observation 
the number of time-independent realisations is limited, which leads to covariance matrices that have partly converged. Moreover, the general case is that it is impossible to acquire the data from the central direction with the truth sensor because of the faintness of the scientific target.
We have therefore developed a method where we fit a model to the off-axis covariance matrices in order to extract the essential parameters of the turbulence and WFS configurations. Therefore, we are able to re-generate the modelled covariance matrices and compute the tomographic estimator from Eq.~\ref{RetrieveTomo}. We call $C_{OffOff\ raw}$ the real on-sky measured off-axis covariance matrix. In the CANARY case during this first run, we also use the TS data to fit the  $C_{OnOff\ raw}$ matrix.
We typically use sets from 10\,000 to 90\,000 synchronised slopes (i.e from $\approx$ 1 to 10~mn at 150~Hz) to compute the on-sky measured covariance matrices $C_{OffOff\ raw}$ and $C_{OnOff\ raw}$. In order to retrieve the parameters $h_l$, $r_0(h_l)$, ($\alpha_i$, $\beta_i$) and ($\alpha_j$, $\beta_j$), we minimize the distance $\epsilon$ between the covariance matrix model to the measured one:
\begin{multline}
\epsilon  =   ||C_{OffOff\ raw} - \sum_l C_{OffOff}\left( h_l, r_0(h_l),    (\alpha_i, \beta_i),   (\alpha_j, \beta_j)\right)||^2 \\
+ ||C_{OnOff\ raw} - \sum_l C_{OnOff}\left( h_l, r_0(h_l), (\alpha_i, \beta_i),   (\alpha_j, \beta_j)\right)||^2\ .
\label{critere}
\end{multline}

We use a Levenberg-Marquardt fitting algorithm to perform the minimisation of $\epsilon$. The number of layers is not retrieved by the algorithm but defined by the user before starting the fitting procedure. This method allowed us to retrieve on-sky the turbulence profile and geometric configuration from the recorded data in a  few minutes. This step is called: \emph{Learn}.
In a sense, the \emph{Learn} resembles a SLODAR analysis, since it allows us to retrieve the $C_n^2(h)$ profile, with the minimization process performing an inversion of the direct problem. However, the $C_n^2(h)$ profile is only a by-product of the {\em Learn}, the real output being the modeled covariance matrices. 
Using the determined turbulence and geometric parameters we compute the theoretical matrices $C_{OffOff}$ and $C_{OnOff}$, then using Eq.~\ref{RetrieveTomo} we compute the tomographic estimator $M_t$ which will be used in the RTC. We called this step \emph{Apply}.

To retrieve the parameters we used the following procedure:
A first calibration is ran on the bench using one phase screen to simulate the Ground layer. We record a set of WFS slopes on all the WFS. The parameters fitted (using LM algorithm) are the pupil misalignments (x,y, theta and  magnification G) in addition to the altitude of the layer (known with a limited optical precision on the bench). Once the pupil misalignments known we use them to align better the system and iterate until the x,y, theta and G are close to 0 (or 1 for G) and become negligible.
Simultaneously we fit the strengths and altitudes of the layers together with the on-sky WFS positions (see section \ref{profile}).


\subsection{Software implementation.}
\label{implementation}
 
In MOAO mode, $M_t$ is an output of the L\&A tomographic algorithm. We emphasise that this matrix is able to reconstruct the slopes that the central sensor should see ($\vec{S_c}$) from the correlation with the off-axis slope measurements $\left( \vec{S_1}, ..., \vec{S_n} \right)$. Therefore, $M_t$ is a \emph{slope tomographic estimator} for the central WFS:
\begin{equation}
\vec{S_c} = M_t \left(
\begin{array}{l}
	\vec{S}_1\\
	 \cdots\\ 
	 \vec{S}_{n}
	 \end{array}
	 \right)\ .
\end{equation}
In the phase A configuration of CANARY we have 3 off-axis WFS and 1 central WFS (TS). For practical implementation we also consider the TS as a fourth off-axis WFS and we have
\begin{equation}
\vec{S_c} = M_t  \left(
\begin{array}{l}
	\vec{S}_1\\
	\vec{S}_2\\
	\vec{S}_3\\
	\vec{S}_{TS}
	 \end{array}
	 \right)\ .
\end{equation}

The three left quarters of the matrix $M_t$, corresponding to off-axis parts, are computed according to Eq.~\ref{equationTomoSol}, the right quarter is zero-padded in order to ignore the TS slopes in the command.
We can therefore rewrite $M_t$ as
   \begin{equation}
   M_t =    \left(
     \begin{array}{ccccccccccccccccccccc}
	M_{t1} & M_{t2}   & M_{t3}   & 0\\
     \end{array}
     \right)\ .
    \end{equation}
  
   In GLAO mode, we average the slopes from the 3 off-axis WFSs leading to the matrix $M_t{_{GLAO}}$ defined as:
   \begin{equation}
   M_t{_{GLAO}} =    \frac{1}{3} \left(
     \begin{array}{ccccccccccccccccccccc }
	Id & Id  & Id  & 0\\
     \end{array}
     \right) 
    \end{equation}
  with $Id$ being a square identity matrix.
The GLAO scheme presented here is achieved with the DM operating in open-loop. This is slightly different to the usual GLAO definition that implies closed-loop operation. The term GLAO used in the rest of this paper describes our averaging of the open-loop off-axis WFS measurements, hence open-loop GLAO.

Finally, as the TS is placed after the DM, we can use it to close the loop in SCAO mode. We disable the off-axis WFS slopes by filling the off-axis WFS part of the reconstructor with zeros. The SCAO reconstructor becomes
  \begin{equation}
   \label{mctSCAO}
   M_t{_{SCAO}} =    \left(
     \begin{array}{ccccccccccccccccccccc}
	0 & 0  & 0  & Id \\
     \end{array}
     \right)\ .
    \end{equation}
   
In CANARY, the TS command matrix $M_c$  (see also Sect.~\ref{interactionmatrix}) is a 72-by-54 array. This command matrix is used to link the slopes in the central direction ($\vec{S_c} $, for instance predicted by the $M_t$ matrix) to the voltage to apply on the DM. The final tomographic command matrix (the one loaded in the RTC), is noted $M_{ct}$ and is defined by
  \begin{equation}
M_{ct} = M_c M_t\ .
\end{equation}
  Depending on the mode we are running (MOAO, GLAO or SCAO), we use the corresponding $M_t$ matrix ($M_t$, $M_t{_{GLAO}}$ or $ M_t{_{SCAO}}$).
  
\subsection{AO controller}
\label{Control}
In close loop configuration (SCAO) we use a conventional integrator temporal controller which takes the form
 \begin{equation}
  \label{closeloopEq}
\vec{V}_{t} = \vec{V}_{t-1} + g.M_{ctSCAO}.
\left(
\begin{array}{l}
	\vec{S}_1(t)\\
	 \vec{S}_2(t)\\ 
	 \vec{S}_3(t)\\ 
	 \vec{S}_{TS}(t)
	 \end{array}
	 \right)
\end{equation}
with $\vec{S}_{i}(t)$  the current slopes vector measurement and $\vec{S}_{ref\ i}$ the reference slopes vector of the $i^{\mathrm{th}}$ WFS. The parameter $g$ is the temporal loop gain of the integrator \citep{Gendron95}. In the RTC implementation, $g$ is a vector with a size of the number of actuators (54). It is possible in particular to filter  the DM and the Tip-Tilt voltages differently. Notice in Eq.~\ref{closeloopEq} that the measurements from the off-axis WFSs are unused due to the zero terms in the identity matrix defining $M_{ctSCAO}$ (Eq.~\ref{mctSCAO}). Only the slopes from the TS are actually used to compute the DM voltages in this close loop configuration. 

The open-loop controller is a temporal filter of the form
  \begin{equation}
  \label{openloopEq}
\vec{V}_{t} = (1-g)\vec{V}_{t-1} + g.M_{ct}
\left(
\begin{array}{l}
	\vec{S}_1(t)\\
	 \vec{S}_2(t)\\ 
	 \vec{S}_3(t)\\ 
	 \vec{S}_{TS}(t)
	 \end{array}
	 \right)\ .
\end{equation}
This controller is used for both GLAO and MOAO open loop modes. As the last quarter part of the $M_{ct}$ is filled with zeros, only the off-axis slopes are used to compute the vector $\vec{V}_{t}$. 

The final set of voltages to be applied to the DM by the RTC is in fact the vector $\vec{U}$: 
\begin{equation}
\label{staticOffset}
\vec{U}(t) = \vec{V}(t) + \vec{V}_{offset}\ .
\end{equation}
The offset vector $\vec{V}_{offset}$ is the static shape to be applied to the DM in order to produce the best static point spread function (PSF) on the IR camera. The computation of this vector is given in the next section. 

\subsection{Reference slopes calibration and DM offset}
\label{RefslopesCal}
The TS reference slopes $\vec{S}_{refTS}$ are deduced by recording the average position of the SH spot, $\vec{S}_{PlaneWF}$, with a reference source producing a flat wavefront placed at the entrance of the TS. We then close the loop on the TS in SCAO mode on the internal calibration source. 
We measure the Non Common Path Aberrations (NCPA) between the TS iteratively with a phase diversity algorithm \citep[see][]{Sauvage2007}. We subtract their contribution, expressed in terms of slopes $\vec{S}_{NCPA}$, to the plane wavefront reference slopes to compute the final TS reference slopes: $\vec{S}_{refTS} = \vec{S}_{PlaneWF} - \vec{S}_{NCPA}$. 

The determination of $\vec{S}_{ref\ i}$ is different.
In open loop, the off-axis WFSs measure their own aberrations, plus the telescope field aberrations and the derotator quasi-static aberrations in addition to the turbulence. 
These static aberrations are not compatible with the tomographic model underlying  the estimator $M_t$, and they must be treated separately. The strategy is just to consider them as constant over the interval of observation, and subtract them from the measurements as reference slopes.
We measure them \emph{in-situ} by averaging a long time sequence of the turbulence (we use the same sequence that served to compute the tomographic estimator) measured by all the WFSs: $\vec{S}_{ref} =  <\vec{S}_{turbu}>_t +  \vec{S}_{static}$.
As the turbulence has a zero average $<\vec{S}_{turbu}>_t= 0$, the average value of slopes measured by the off-axis WFSs are their corresponding static aberrations in the field and are set as reference slopes $\vec{S}_{ref\ i} =  \vec{S}_{static\ i}$.

\subsection{DM offset voltages}
\label{DmOffsetVoltage}
The on-axis telescope and derotator quasi-static aberrations $\vec{S}_{static\ TS}$ can  also be measured using the same procedure. These measured on-axis aberrations are subtracted from the TS reference slopes in order to find the aberration slopes. To ensure their compensation on-axis, these slopes have to be converted into a static voltage vector $\vec{V}_{offset}$ (also taking into account the NCPA):  
\begin{equation}
\label{offset}
\vec{V}_{offset} = - M_c (\vec{S}_{static\ TS} - \vec{S}_{PlaneWF} + \vec{S}_{NCPA})\ .
\end{equation}
The vector $\vec{V}_{offset}$ remains applied on the DM (Eq.~\ref{staticOffset}) during the whole duration of the observation in open loop. 

This strategy works provided the condition $<\vec{S}_{turbu}>_t= 0$ is respected, otherwise a static error will propagate through the estimator $M_t$ due to incorrect reference slopes, and will add to the DM due to an incorrect $\vec{V}_{offset}$. This is exactly what is observed in reality, but we are fortunately able to demonstrate that those two errors terms opportunely cancel each other (or at least partly, under certain conditions). For the sake of simplicity in this explanation, we will lighten our notation. We call $s$ the vector of the static term for the off-axis WFS, and $t$ the turbulence. We write $s'$ and $t'$ the same quantities measured by the on-axis truth sensor. We also define $r$, the reference slope vector on the truth sensor. As said previously, the average of the quantities related to turbulence are non-zero, and we have $<t>=\epsilon_t \ne 0$ and $<t'>=\epsilon_t'\ne 0$. Our hypothesis is that, as the estimator $M_t$ is linear and designed to work on turbulence, we have 
\begin{equation}
\label{linearavg}
\left\{
\begin{array}{l}
M_t t = t' \\
M_t \epsilon_t = \epsilon_t'\ .
\end{array}
\right.
\end{equation}
Now, at the end of the calibration of the reference slopes and the DM offsets, the off-axis WFS will be set with $(s+\epsilon_t)$ as their reference slopes, and the DM offset will be $-(s'+\epsilon_t'-r)$ (we skip $M_c$ for the sake of simplicity). Both terms contain errors, as they include partly-converged quantities $\epsilon$. However, now the measurements from the off-axis WFS for any turbulence $t$ are 
$(s + t - s - \epsilon_t)$. Thus, the total voltage applied on the DM will be $-M_t(t-\epsilon_t) -(s'+\epsilon_t'-r)$, and the wavefront on-axis will be $s'+t'-M_t(t-\epsilon_t) - (s'+\epsilon_t'-r)$. Taking into account relation~\ref{linearavg}, it follows that all terms vanish and the expression is left with $r$. The TS will measure the wavefront with respect to its own reference slopes $r$, which gives $r-r=0$, i.e. a flat, corrected wavefront, in spite of the errors in both the reference slopes and the DM offset voltages.

The success of the on-sky measurement method of all static offsets relies on the fact that 
$\vec{S}_{static\ i}$ and $\vec{V}_{offset}$ are estimated from the {\em same} set of data. In this case, the error we make is related only to partial convergence of the average of non-tomographic aberrations (unseen layers, drifts in the instrument or telescope, etc) and not to the partial convergence of the average of the turbulence itself.

\section{Data reduction}
\label{reduc}
We have PSF images on the target direction (i.e. the central star) taken using the IR camera. In parallel to the IR images, the real-time slopes of all WFSs 
(off-axis and TS) plus DM voltage data are saved to determine the atmospheric parameters and evaluate the error budget. We record two types of synchronous data for all the WFSs:
\begin{itemize}
\item
\emph{engaged slopes} recorded while the MOAO loop is engaged. It stands for the open loop sensing on the 3 off-axis WFS (used to compute the correction on-axis) and residual slopes seen by the TS. 
\item 
\emph{disengaged slopes} recorded while the MOAO loop is NOT running. DM is flattened allowing measurements of the turbulence in open loop on each of the 4 NGS WFS directions.
\end{itemize}
We now present  the procedure used to reconstruct different terms in the error budget  of CANARY from these data.

\subsection{IR image performance estimation}
\label{IRcalcul}
The background is subtracted from the IR images and dead pixels are removed using a pixel map previously established
on a dark image. The IR images shown here are simply the average of 30 individual exposures of 1 second (not tip-tilt removed). The IR image was centered in a region of the detector with no dead pixel to avoid any bias in the estimation on the PSF.

Strehl ratios (SR) have been computed on the IR images by normalizing their total energy to unity, and dividing their peak
value by that of the diffraction-limited pattern sampled identically. This peak value is given by $a=\frac{\pi}{4}(D^2-o^2)p^2/\lambda^2$ with $o$ the central obscuration diameter, and $p$ the camera pixel scale in rd~pixel$^{-1}$. 
First, we calibrate the pixel scale using an internal source at $\lambda=$1550~nm, and we ensure that the WFS pixel scale and IR camera pixel scale are consistent together. This is important to ensure that all absolute WFS-related values (r0, and any nm rms value measured from wavefront data) will be consistent with the IR camera pixel scale. Then a supplementary calibration has been done on-sky using a reference double star imaged on the IR camera to determine the final pixel scale.
We close the loop using the TS and measure the Optical Transfert Function (OTF) of the IR image. From the OTF, we estimate the cut-off frequency at D/$\lambda$ and check the number of pixels in $\lambda/D$.
The normalisation of the flux is dramatically sensitive to the estimation of background level to be subtracted. Special care has been taken with this operation, first narrowing the field to only 64$\times$64 pixels around the source and then estimating the residual background level using edge pixels. We estimate the uncertainty on the SR due to normalization errors in the background to be of the order of 0.02 (2\%). 

\subsection{Seeing estimation}
\label{seeing}
The off-axis WFS continuously measure the open-loop turbulence, irrespectively of the loop being engaged, when operating in MOAO or SCAO modes. They can always be used to estimate the Fried parameter $r_0$ at any time.
For each of the 3 off-axis WFS, $r_0$ is computed by fitting the theoretical variances of the Zernike decomposition of the Kolmogorov spectrum~\citep{Noll76} to those of the experimentally reconstructed wave-front.
Before fitting, the experimental variances are corrected from the wave-front sensing noise measured on the slopes as explained in Sect.~\ref{Noise}, and propagated onto the Zernike coefficients. The wave-front is reconstructed on Zernike modes from $Z_2$ to $Z_{36}$, but the fit only takes into account modes 4 to 27.
Tip-tilt is excluded, as it may be polluted by the telescope tracking or vibrations, and is definitely influenced by the outer scale $L_0$. We did not observed any significant vibration on the temporal spectra of the higher order modes. Modes 28 to 36 are excluded too, because as they are the last radial order they are more affected by aliasing effects.
$r_0$ is given at $500$ nm and at the airmass of the observation (not rescaled to zenith). The final $r_0$ estimation seen by CANARY is computed by taking the mean of the estimated $r_0$  from the 3 off-axis WFS measurements.

\subsection{CANARY error budget}
\label{CANARYErrBudget}
The estimated wavefront error $\sigma_{Err}$ can be translated in an expected SR using the formula SR = exp(-(2$\pi\sigma_{Err}/ \lambda)^2)$ and compared to the  SR measured on the IR image.
The overall error budget of CANARY is given for the IR on-axis channel where the images are recorded. 
The total error budget on the IR camera, denoted $\sigma_{ErrIR}$, can be expressed as
\begin{multline}
\label{IRTotal}
\sigma^2_{ErrIR} =  \sigma^2_{NCPA} + \sigma^2_{FieldStat} +  \sigma^2_{Fit} + \sigma^2_{TomoNoiseFilt}  \\
+ \sigma^2_{BW} + \sigma^2_{AliasGround} + \sigma^2_{AliasAlt} + \sigma^2_{Tomo} + \sigma^2_{OL}\ ,
\end{multline}
and we will assume in this paper that all these terms are independent, so that variances add up together. We define now each of the individual terms of the error budget and describe how we compute them. 

%

\subsubsection{NCPA error $\sigma^2_{NCPA}$}
\label{NCPA}

The best SR obtained on-bench, after the NCPA estimation by phase diversity and compensation by the DM, is 0.80 $\pm$ 0.02 ($\approx$ 115~nm rms ). It corresponds to residual NCPA and the high spatial frequencies non correctable by the DM. We call the both contributions as NCPA noted $\sigma_{NCPA}$.  The best flat of the DM is 50~nm rms determined from interferometric measurements measured in May 2009. 
The poor error figure is explained by high spatial frequencies that developed between actuators due to the aging of the mirror.

\subsubsection{Field static aberrations $\sigma^2_{FieldStat}$}
In MOAO, the off-axis WFSs and the TS measure, in addition to the atmospheric turbulence, static aberrations across the telescope field of view. For the TS the DM creep is also included in this term. These aberrations mainly come from the telescope and the derotator and may vary across the field. We consider them as quasi static aberrations as they slowly evolve during the night. We calibrate and subtract the field aberrations in the MOAO loop. However, a non-perfect determination of the off-axis static aberrations leads to a residual static error after the MOAO correction in the on-axis direction. Any additional DM creep due to the static term $V_{offset}$ applied to the DM during open-loop operation can also contribute to this error term.
The total on-axis static error noted $\sigma_{FieldStat}$ can be measured by the TS by averaging its slope measurements while the loop is engaged.

This error term is related to the inaccuracies in the calibration of the static offsets, that have already been described in Sect.~\ref{DmOffsetVoltage}.

\subsubsection{Fitting error $\sigma^2_{Fit}$}
The estimation of the DM fitting error $\sigma_{Fit}$ (or undermodelling error) has been determined using a Monte-Carlo simulation. We have computed the residual phase variance of computer-generated Kolmogorov wave-fronts subtracted from their best least-squares fit on the DM modes. We have restricted the modes to the 47 actually controlled by the system, out of the 54 degrees of freedom (52 on the DM, 2 on the tip-tilt mirror). The 7 filtered modes correspond to a resulting conditioning number of 50.
This fitting error was found to be 
\begin{equation}
\label{fitcanary}
\sigma^2_{Fit} = 0.0122 \left( \frac{D}{r_0} \right) ^{5/3}.
\end{equation}
Then, the fitting error term of our error budget is a number which will simply be derived from the knowledge of $r_0$, using the above equation.

\begin{figure}
\centering
\includegraphics[width=5cm]{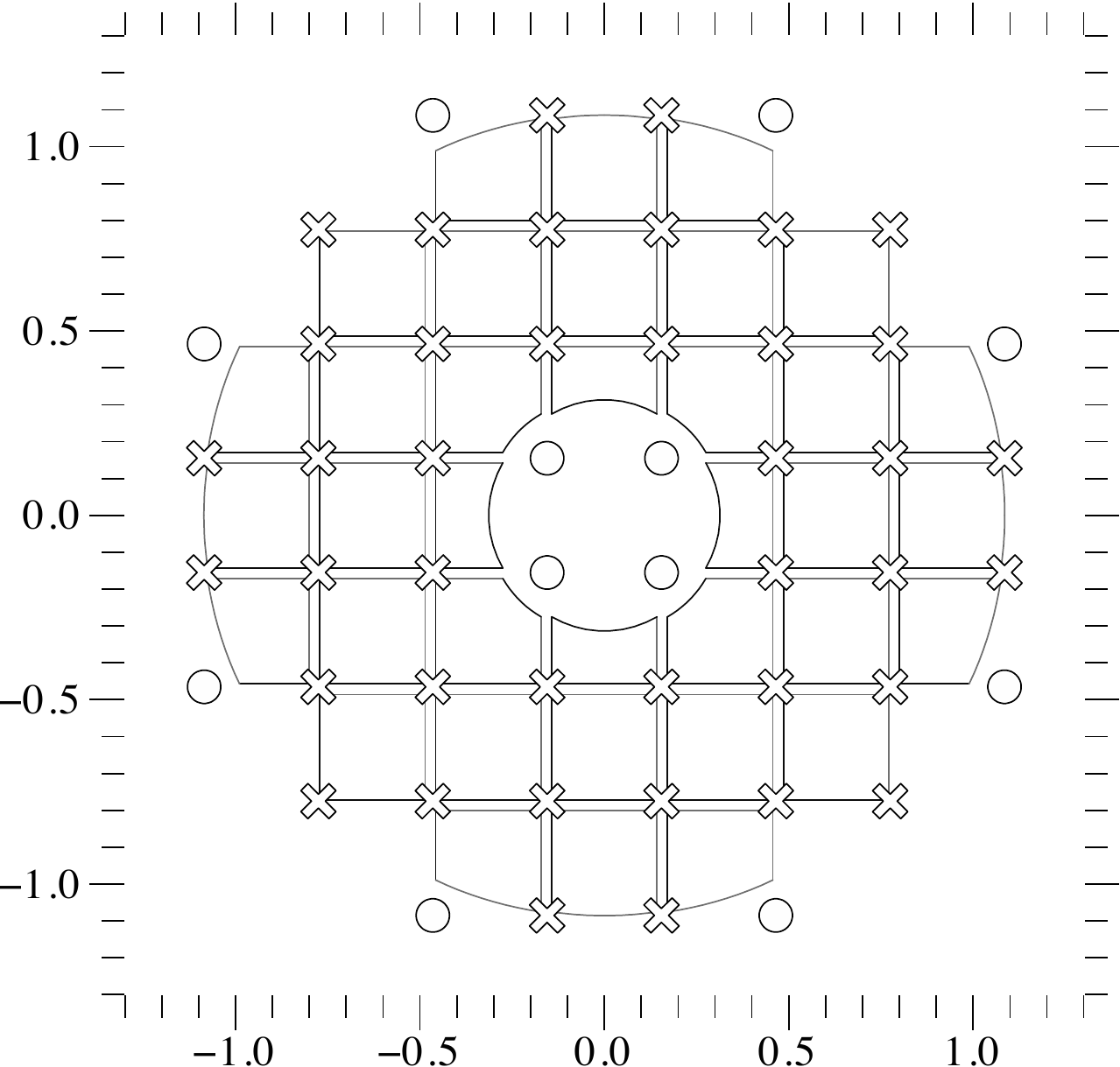}
\caption{Geometric configuration of the 36 useful sub-apertures of a Shack-Hartmann WFS of CANARY across the WHT pupil (with a 0.286 normalised diameter central obscuration). Actuators within the pupil are represented with a cross. Actuators out of the pupil are also represented with circles.}
\label{pupil}
\end{figure}

\citep{Roddier1999} gives the fitting error as function of the number of actuators $n_a$ in the pupil diameter:
$\sigma^2_{Fit} = 0.335 n_{a}^{-5/3} ( {D}/{r_0} ) ^{5/3}$ 
and
replacing $n_a$ with $2 \sqrt{N_{tot}/\pi}$ leads to
\begin{equation}
\sigma^2_{Fit} = 0.274  \, N_{tot}^{-5/6} \left( \frac{D}{r_0} \right) ^{5/3},
\end{equation}
which gives $N_{tot}=42$ in our case when identifying with Eq.~\ref{fitcanary}. This is extremely close to the number of actuators that are truly laying within the pupil, as depicted on Fig.~\ref{pupil}. This latter represents the configuration of the subapertures within the pupil and of the actuators of CANARY. One can see that a total of 12 actuators (represented with circles) are poorly seen by the WFS, leading to only 40 actuators in the pupil, a value compatible with $N_{tot}=42$.

Now, we can also compare Eq.~\ref{fitcanary} with this one derived by \citep{Conan94},
\begin{equation}
\label{Noll}
\sigma^2_{Fit} = 0.257 \, N_{Z}^{-5/6} \left( \frac{D}{r_0} \right) ^{5/3}
\end{equation}
that gives the fitting error after the correction of the $N_z$ first Zernike polynomials. We find that our DM amounts to $N_Z \approx 38$ Zernike modes.

In the following sections of this article, we express all error terms in~nm~rms. They are computed by using a Zernike decomposition of the WFS slopes. We have chosen to consider a wavefront reconstruction on $N_Z$= 36 Zernike polynomials corresponding to radial order $n=7$, as this is nearly equivalent to our DM.
We define $M_{iz}$ the matrix containing the response of the SH to the Zernike modes 2 to 36, and $M_{rz}$ the Zernike reconstruction matrix (generalized inverse of the $M_{iz}$ matrix).

\subsubsection{Noise error $\sigma^2_{Noise}$}
\label{Noise}
We compute the  noise $\sigma^2_{Noise}$ on the WFS slope signals from their temporal autocorrelation. 
As noise and turbulence are two independent processes, the autocorrelation of their sum is the sum of autocorrelations.
At $\Delta t$=0, in particular, the measured variance is the sum of variances of noise and turbulence. 
While the autocorrelation of noise is a Dirac function, that of turbulence is wide and locally smooth around $\Delta t =0$ : this allows us to separate them by fitting a parabola $(a . \Delta t^2+b)$ to the two points of the autocorrelation at $\Delta t$=1 and $\Delta t$=2 and extrapolating the purely turbulent variance at $\Delta t = 0$. This value is subtracted from the total variance at $\Delta t = 0$, to obtain the noise variance on each slope. The WFS slope noise is then propagated through the Zernike reconstruction matrix $M_{rz}$ on 36 modes in order to be converted into a wavefront error. We compute the propagated noise by making the assumption that there is no spatial correlation between noise on any of the slopes, i.e. the covariance matrix of the slope noise is a pure diagonal matrix, noted $\sigma_{slopesnoise}^2$. The propagated noise is then given by
\begin{equation}
\label{noisePropagation}
\sigma^2_{Noise} = \mathrm{Trace}(M_{rz} Diag(\sigma_{slopesnoise}^2) M_{rz}^t) \ .
\end{equation}
This equation is used for the TS noise computation  $\sigma^2_{NoiseTS}$. The diagonal terms on the noise matrix depends on the GS magnitude and considered sub-aperture.

\subsubsection{Noise through the tomographic estimator $\sigma^2_{TomoNoiseFilt}$}
\label{noiseTomo}
This term is computed from noise on the off-axis WFS, computed as described in the previous paragraph, propagated through the tomographic slopes estimator matrix $M_t$, and propagated through the Zernike reconstruction matrix $M_{rz}$: 
\begin{equation}
\sigma^2_{TomoNoise} = \mathrm{Trace}\left( (M_{rz}M_t) Diag(\sigma^2_{Off-axis\ slopesnoise}) (M_{rz}M_t)^t  \right)\ .
\end{equation}
This white noise is filtered by the temporal controller (see Sect.~\ref{Control}, Eq.~\ref{openloopEq}), and only a fraction of it is actually injected in the DM command. The reduction of the noise variance due to the low-pass filtering is given by a factor of $\frac{g}{2-g}$ (demonstrated in Appendix~\ref{append_g/2-g}), where $g$ is the loop gain:
\begin{equation}
\label{TomoNoisefilt}
 \sigma^2_{TomoNoiseFilt} = \left( \frac{g}{2-g} \right) \, \sigma^2_{TomoNoise}\ .
\end{equation}

\subsubsection{Bandwidth error $\sigma^2_{BW}$}
\label{bandwidthError}
The bandwidth error is computed using a set of recorded engaged-loop WFS data using the off-axis WFS with the highest signal-to-noise ratio (in open loop).
The principle is to numerically filter this real data set in order to reproduce frame-by-frame the loop behaviour in engaged mode, then derive the error from the simulated residuals. The method ensures 
the computation of a bandwidth error term that corresponds to the precise recorded turbulence conditions.

A first step computes the open loop controller filtering (see Sect.~\ref{Control}) of the slope measurements:
\begin{equation}
\label{filtFirstOrder}
S'_{t} = (1 - g) \, S_{t-1} + g \, S_{t}\ .
\end{equation}
Then a second step simulates the application of the DM command, which acts as a zero-order hold, and that occurs partly during the integration of the subsequent frame $t+1$, and partly during the frame after that at  $t+2$. Writing the fractional delay, expressed in frames, as $(1+\alpha)$, with $0 \le \alpha < 1$,
then the residual signal $R_t$ is
\begin{equation}
\label{decalFraction}
R_t = S_t - (\alpha \, S'_{t-2} + (1-\alpha) \, S'_{t-1}) \ ,
\end{equation}
and the variance of this last signal $R_t$ characterizes the bandwidth error, except it just needs to be unbiased from noise.

As the on-sky data $S_t$ unavoidably includes noise, we need to correct the variance of $R_t$ from the noise variance that has propagated through our filtering process, since we aim to compute here the bandwidth error associated only with turbulence (the impact of noise has been treated in ~\ref{noiseTomo}).
It is demonstrated in Appendix~\ref{append_gaga} that a pure white noise of unitary variance, filtered as described in Eqs.~\ref{filtFirstOrder} and~\ref{decalFraction} is output with a variance given by $(1 - 2 g \alpha (1-\alpha))g/(2-g)$.

Finally, to express the bandwidth error as a wavefront error, we reconstruct the measurements $R_t $ in a Zernike basis using the matrix $M_{rz}$, and we compute the variance of each Zernike coefficient.
The noise variance on each slope is also propagated on the Zernike coefficients as described in Eq.~\ref{noisePropagation}, the factor $(1 - 2 g \alpha (1-\alpha))g/(2-g)$ is applied on each, and the result is subtracted from the variance of $R_t $. The BW error term also includes the vibration error term since the integrator controller does not efficiently reject the vibrations. The vibration term in CANARY has been studied in \citep{Kulcsar2012}. We measured Tip-Tilt vibration peaks between 20Hz and 50Hz and of the order of 100nm rms for the night of the 27th of September.

\subsubsection{Aliasing error $\sigma^2_{Alias}$}
\label{aliasing}
Aliasing is due to the fact that the WFS spatially samples the wavefront across the pupil of the instrument. The high spatial frequencies of the wavefront disturbances will be undersampled if they are higher than the Nyquist frequency (i.e. higher than the half of the sampling frequency). Thereby, these high frequencies are mistaken for low spatial ones. Another way to say it is that the incoming wavefront $\phi$ does not entirely lie within the mirror subspace, it contains a component $\phi_\perp$ orthogonal to it (thus giving rise to the fitting error). This orthogonal wavefront is seen by the wavefront sensor but unfortunately produces non-zero measurements that will be reconstructed and mistaken by the system as mirror modes. 
\citep{quiros2010_genAliasing} has given an extensive analysis of aliasing in tomographic applications, and we do not aim to do the same here. We will use a simplified, approximate approach to disentangle the different aliasing effects on the system. As already stated in section \ref{CANARYErrBudget} our main assumption is a negligible coupling between the terms of the error budget, in particular $\sigma^2_{Alias}$ with $\sigma^2_{BW}$ and $\sigma^2_{Tomo}$.

First, we compute the covariance matrix of the turbulence on the Zernike basis $C_{zz}$ with 900 polynomials (which we assume to be large enough to represent an infinite number) normalized to $D/r_0=1$  using the formulae given by \citep{Noll76}. Then, we 
zero the lines and rows corresponding to the first modes up to $Z_{36}$ in order to obtain a
covariance matrix of the turbulence $C_{zz\perp}$ only of high orders (of $\phi_\perp$, as named above).  
This matrix is supposed to mimic the statistics of the phase orthogonal to the DM space.
We then compute the slopes covariance matrix of this high-order turbulence, noted $C_{ss}$, using a transformation matrix from Zernike to slopes $M_{iz\ 900}$, computed here for the first 900 Zernike polynomials:
\begin{equation}
C_{ss} = M_{iz\ 900} \, C_{zz\perp} \, M_{iz\ 900}^t\ .
\end{equation}
The matrix $C_{ss}$ represents the statistics of the aliased wavefront on the WFS.
We finally use the Zernike reconstruction matrix $M_{rz}$ on the $C_{ss}$ slopes covariances matrix to compute the aliasing error,
\begin{equation}
\label{aliasGendron}
 \sigma^2_{Alias} = 
 \mathrm{Trace}\left( M_{rz} C_{ss}  M_{rz}^t \right) \  \left( \frac{D}{r_0} \right)^{5/3} \, ,
 \end{equation}
and we should now evaluate how the aliasing will propagate through the control. For this, we consider the aliasing effect differently depending on whether layers are placed in altitude or at the ground.

We assume that aliasing produced by layers located close to ground level is fully correlated between all the WFSs including the TS. This aliasing error is injected into the loop and is consequently fully applied to the wavefront by the DM. This introduces an error on the IR camera that is not seen on the TS slopes in first approximation. We compute its effects as a fraction $X_{Ground}$ of the total turbulence on the telescope:
\begin{equation}
 \sigma^2_{AliasGround} = X^{}_{Ground} \   \sigma^2_{Alias}\ .
 \end{equation}

Contrarily to ground, we assume the aliasing produced in altitude behaves as a white spatial noise, not correlated between the off-axis WFSs. We consider to be a noise contribution by averaging on the  WFSs,
\begin{equation}
\sigma^2_{AliasAlt} = X^{}_{Altitude} \  \frac{\sigma^2_{Alias}}{n_{WFS}}\ ,
 \end{equation}
where $ X_{Altitude}$ is the fraction of the turbulence which is not at the ground and $n_{WFS}=3$ the number of WFSs. Using the last two equations, we neglect the impact of the open loop filtering on the aliasing terms. 
Finally, the aliasing in altitude, produced only by the TS, is computed by
\begin{equation}
\sigma^2_{AliasAltTS} = X^{}_{Altitude} \   \sigma^2_{Alias}\ .
\end{equation}

\subsubsection{Tomographic error $\sigma^2_{Tomo}$}
\label{tomo}

We compute the tomographic error on a given data set by estimating the residuals between the non-engaged measurements of the TS and the synchronous tomographic prediction. We also have to unbias the tomographic error from several additional effects, that we detail in Eq.~\ref{tomoeq}. This latter is computed by multiplying the off-axis slopes $\vec{S}_{offAxis}$ of the set with the tomographic estimator $M_t$.

When we are dealing with engaged-loop data, we do not have any non-engaged measurements of the TS.
Our first attempt for retrieving non-engaged TS data was to subtract the contribution of the DM (with a proper multiplication with the interaction matrix $M_i$, Sect.~\ref{interactionmatrix}). This approach is incorrect, because it will include the DM open-loop error that cannot be disentangled from a tomographic error.
We therefore decided to instead use a set of slopes taken with the loop \emph{disengaged}, just before or after the engaged-loop set, to determine the tomographic error of the latter. This method works provided the value of $\sigma^2_{Tomo}$ is properly rescaled with respect to the $r_0$ value. A factor of $(r_0/r'_0)^{5/3}$ has to be applied to the variance.


We then estimated the raw tomographic error by computing the difference between this tomographic prediction and the real measurements made by the TS without introducing any delay for each frame. We express the vector of the residuals onto a Zernike basis:
\begin{equation}
\{ a_{i\ Tomoraw}  \}  = M_{rz}(\vec{S}_{TS} - M_t \vec{S}_{offaxis})\ .
\end{equation}
The reference slopes are already subtracted from $\vec{S}_{offaxis}$ and $\vec{S}_{TS}$ as described in section \ref{LandA}. Then the raw tomographic error $\sigma^2_{Tomoraw}$ is simply given by the sum of the variances of Zernike coefficients:
\begin{equation}
\sigma^2_{Tomoraw} = \sum^{36}_{i=2} Var\left( \{  a_{i\ Tomoraw} \} \right)\ .
\end{equation}

As we compare perfectly synchronised disengaged slope data, we have a direct access to the tomographic error with no temporal effect from the loop filter. However, it still needs to be corrected from the impact of noise and aliasing.
We remove the TS noise  $\sigma^2_{NoiseTS}$ and the propagated noise from the off-axis WFSs $\sigma^2_{TomoNoise}$. For aliasing, we remove the contribution of  the aliasing effect in altitude on the TS $\sigma^2_{AliasAltTS} $ and propagated from the off-axis WFSs $\sigma^2_{AliasAlt}$.
It follows that, provided \textit{disengaged slopes}, the pure tomographic error $\sigma^2_{Tomo}$ can be estimated from
\begin{multline}
\label{tomoeq}
\sigma^2_{Tomo} = \sigma^2_{Tomoraw} - \sigma^2_{TomoNoise} - \sigma^2_{NoiseTS} \\
- \sigma^2_{AliasAltTS} -\sigma^2_{AliasAlt} \ .
\end{multline}

We emphasize that the on-sky measured term $\sigma^2_{Tomo}$ could itself be split
into different terms, in particular the error term $\sigma^2_{Model}$ of the turbulence profile model, corresponding to an error made on the model used to build the reconstructor. 
We can have an estimation of this model error by comparing the $\sigma^2_{Tomo}$, evaluated on-sky with the tomographic error evaluated with a numerical simulation using a $C_n^2(h)$ profile strictly equal to the model that served to compute the estimator $M_t$.
We give an example of such a comparison in the last section of this paper.

\subsubsection{Open loop error $\sigma^2_{OL}$}
This term has also been named ``go-to'' error in the literature, and corresponds to the fact that the shape the mirror will take for a given set of voltages is not exactly the one that one would expect, due to hysteresis, drifts, non-linearities or any other effect that has not been taken into account in the mirror model.
We also underline that in our analysis, the term $\sigma^2_{OL}$ only represents the dynamic wave-front error linked to the open loop behavior of CANARY DM. The static wave-front error term is given by 
$\sigma^2_{FieldStat}$. 

As stated in paragraph~\ref{tomo},
the dynamic open loop error is difficult to disentangle from the tomographic error. 
We decide to estimate it from the \emph{engaged slopes} where we can measure the residual wavefront error seen by the TS, denoted hereafter as $ \sigma_{ErrTS}$. The error seen by the TS can be computed as a sum of the individual terms:
\begin{multline}
\label{ErrTS}
\sigma^2_{ErrTS} = \sigma^2_{Tomo} + \sigma^2_{OL} + \sigma^2_{TomoNoiseFilt} + \sigma^2_{NoiseTS} \\
+ \sigma^2_{AliasAltTS} + \sigma^2_{AliasAlt} + \sigma^2_{BW} + \sigma^2_{FieldStat}\ .
\end{multline}
Considering that we measure $\sigma^2_{ErrTS}$, $\sigma^2_{OL}$  is the last unknown in Eq.~\ref{ErrTS}. It can be estimated by
\begin{multline}
\label{ErrOL}
\sigma^2_{OL} = \sigma^2_{ErrTS} - \sigma^2_{Tomo} - \sigma^2_{TomoNoiseFilt} - \sigma^2_{NoiseTS} \\
- \sigma^2_{AliasAltTS} -\sigma^2_{AliasAlt} - \sigma^2_{BW} -  \sigma^2_{FieldStat}\ .
\end{multline}

\emph{Engaged slopes} are therefore required to compute an estimation of the open loop error and also \emph{disengaged slopes} because an estimation of the $\sigma^2_{Tomo}$ term is needed. Because our estimation is computed by subtracting a large number of  estimated terms from $\sigma^2_{ErrTS}$, $\sigma^2_{OL}$ may also include all the estimation errors (finite number of slope samples, approximations, bad calibration, etc). Therefore, it is only a crude estimation of $\sigma^2_{OL}$. 
\section{On-sky Results}
\label{OnskyResults}

We had 2 $\times$ 4 nights split between September (19th, 22nd, 26th and 27th)  and November 2010 (from 23rd to the 26th). Unfortunately due to bad weather we lost most of the November nights.
We focus in this paper on the results obtained on the fourth night of the September run (2010 Sep. 27th). We alternated the observations between the SCAO (closed loop on the TS), GLAO and MOAO modes as the turbulence profile evolved and as we changed asterisms. During the whole night, the sampling frequency of all WFSs was 150~Hz irrespective of the AO mode. Although the temporal controllers are slightly different between open and closed loop, the RTC latency is the same for all modes.

\subsection{Natural guide stars asterisms}
\begin{figure}
\centering
\includegraphics[width=9cm]{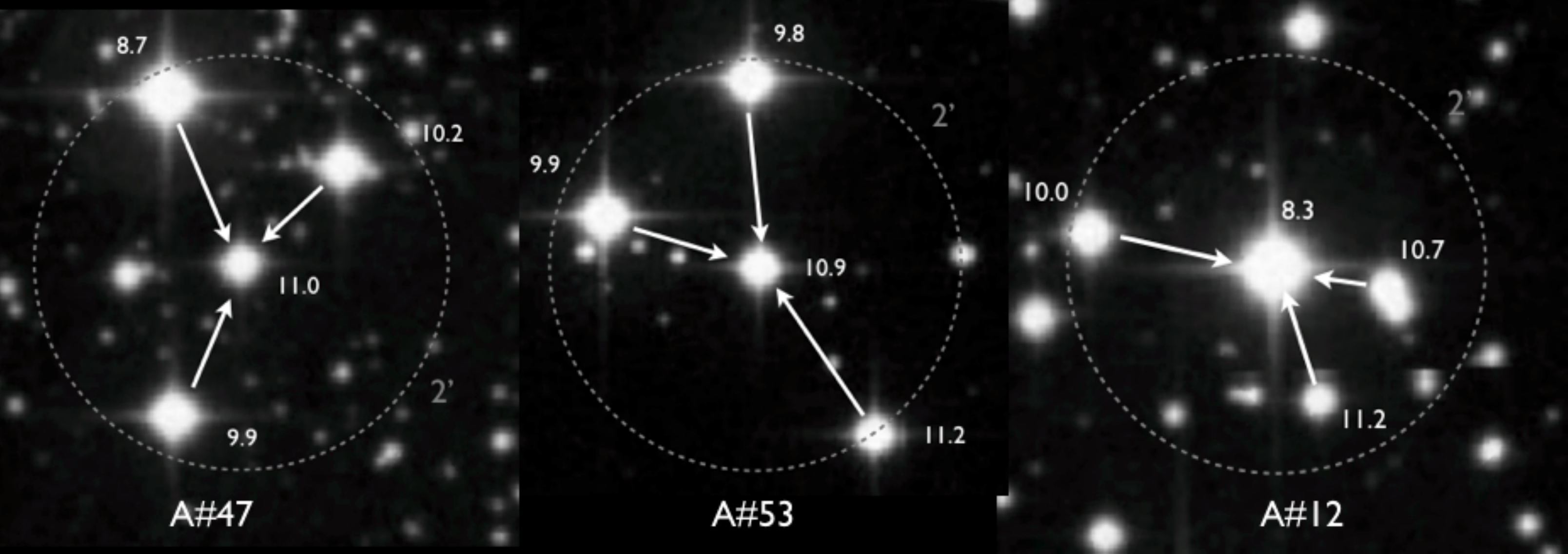}
\caption{Map of the 3 asterisms (DSS images) observed during the September run. The tomographic estimator was computed using the 3 off-axis stars and applied in the central direction (white arrows). The dashed circle represents a 2' diameter field of view.}
\label{Ast}
\end{figure}

  CANARY makes use of star asterisms formed by four NGSs. The central one, placed on-axis and used for diagnostic purposes, mimics the science object that will benefit from the turbulence compensation. We selected asterisms with a distance between the central on-axis star and the three off-axis ones ranging from $15\arcsec$  to $65\arcsec$, while keeping all stars brighter than m$_V$=12. The three observed asterisms of the 54 identified asterisms for the September period are described in Table~\ref{asterisms} and Fig.~\ref{Ast}.

\begin{table}
  \caption[]{Three asterisms observed during the night of the 27th of September 2010. The columns indicate: the CANARY asterism reference number, the separation (in arcsec) of each off-axis star to the central one, the V band magnitudes of each and the range of airmass during observations.}
  \label{asterisms}
  \centering
  \begin{tabular}{llll}
      \hline  \hline
Asterism \#	&	47	&	53	&	12	\\
        \hline
central $m_V$	&	11	&	10.9	&	8.3	\\
        \hline
sep (\arcsec)	&	47.9	&	61.7	&	39.3	\\
$m_V$	&	9.9	&	11.2	&	11.2	\\
        \hline
sep (\arcsec)	&	40.6	&	49.1	&	31.4	\\
$m_V$	&	10.2	&	9.9	&	10.7	\\
        \hline
sep (\arcsec)	&	53	&	56.8	&	51.5	\\
$m_V$	&	8.7	&	9.8	&	10	\\
        \hline
airmass	&	1.02--1.55	&	1.11--1.50	&	1.05--1.09\\
        \hline
  \end{tabular}
\end{table}

\subsection{CANARY tomographic altitude resolution}
\label{tomoreso}
The maximum spatial frequency measured by a WFS is defined by $f_e = \frac{1}{2d_{ssp}}$ with $d_{ssp}$ the size of the sub-aperture projected on-sky ($d_{ssp}$ = 0.6m for CANARY). 
For that frequency, the tomographic altitude resolution $\Delta_h$, is computed with the relation $ \Delta_h =  \frac{d_{ssp}}{\alpha} $, with $\alpha$ the separation angle between 2 WFSs. The maximum altitude accessible $h_{max}$ is given by the relation $h_{max} = \frac{D_{tel}}{\alpha}$. A large separation angle will allow us to take advantage of a good altitude resolution, while small separations give access to a higher sensing altitude.

Because CANARY is equipped with 4 WFSs, each senses the wavefront in a different direction according to the NGS configuration (i.e. the observed asterism). This leads to different tomographic resolutions according to the considered pairs of WFSs. Table~\ref{TomoResolution} summarizes the altitude resolution for each baseline of each asterism. During the night, the altitude resolution was in the range 1\,000 to 4\,000~m. The best resolution of $ \Delta_h=$1\,083~m is achieved with the pairs of WFS 1 and 2 on asterism A53. It also means that we were more sensitive to a variation of a layer altitude on asterism 53 than asterisms A47 and A12. Because of the very large shift in altitude any layer above 7\,500~m cannot be seen by this particular pair of WFS but was completed by the measurement of narrower pairs like WFS 2 and 3 that allowed us to sense the turbulence profile up to 14\,000~m.

\begin{table}
  \caption[]{Vertical altitude resolution and maximum altitude in metres computed for each pairs of the 4 WFSs. There are 6 different combinations computed for each of the 3 observed asterisms. $\alpha$ is the angular separation between the 2 considered stars. }
  \label{TomoResolution}
  \centering
  \begin{tabular}{ l c c c c  }
    \hline \hline 
Asterism        &    star           &  $\alpha$  & $\Delta_h$ &  $h_{max}$  \\
                    &  couple         &        (\arcsec)     & (m)           &        (m)        \\
    \hline
        &  1--C      &  47.9  & 2583   &  18085    \\ 
        &  2--C      &  40.6  &3048   &  21337    \\        
A47   &  3--C      &  53    & 2335   &  16354    \\    
        &  1--2      &  86.5    & 1430   &  10015  \\    
        &  2--3      &  54.8    & 2258   &  15808  \\    
        &  1--3      &  92.4    & 1339   &  9376  \\   
    \hline
        &  1--C      &  62.2  & 1990   &  13927    \\ 
        &  2--C      &  56.7  & 2183   &  15279    \\        
A53   &  3--C      &  48.7    & 2541   &  17789  \\     
        &  1--2      &  114.2    & 1083   &  7586  \\     
        &  2--3      &  60.26    & 2054   &  14376 \\    
        &  1--3      &  105.27    & 1176   &  8229 \\    
    \hline
        &  1--C      &  38.1  & 3248   &  22738    \\ 
       &  2--C      &  31  & 3992   &  27946      \\        
A12  &  3--C      &  51.8    & 2389   &  16724  \\    
        &  1--2      &  35.5    & 3486   &  24403  \\    
        &  2--3      &  82.9    & 1493   &  10450  \\    
        &  1--3      &  77.9    & 1589   &  11121  \\    
    \hline
  \end{tabular}
\end{table}

The following sections present the evolution of the seeing conditions, the SR measured on the IR camera, the turbulence profile, the tomographic error, the open loop error and the residual field aberrations during the night of September 27th.
Section~\ref{details} presents a detailed analysis of the error budget for three different cases during the night.

\subsection{Seeing conditions}
\label{evolution}
\begin{figure}
\centering
\includegraphics[width=8cm]{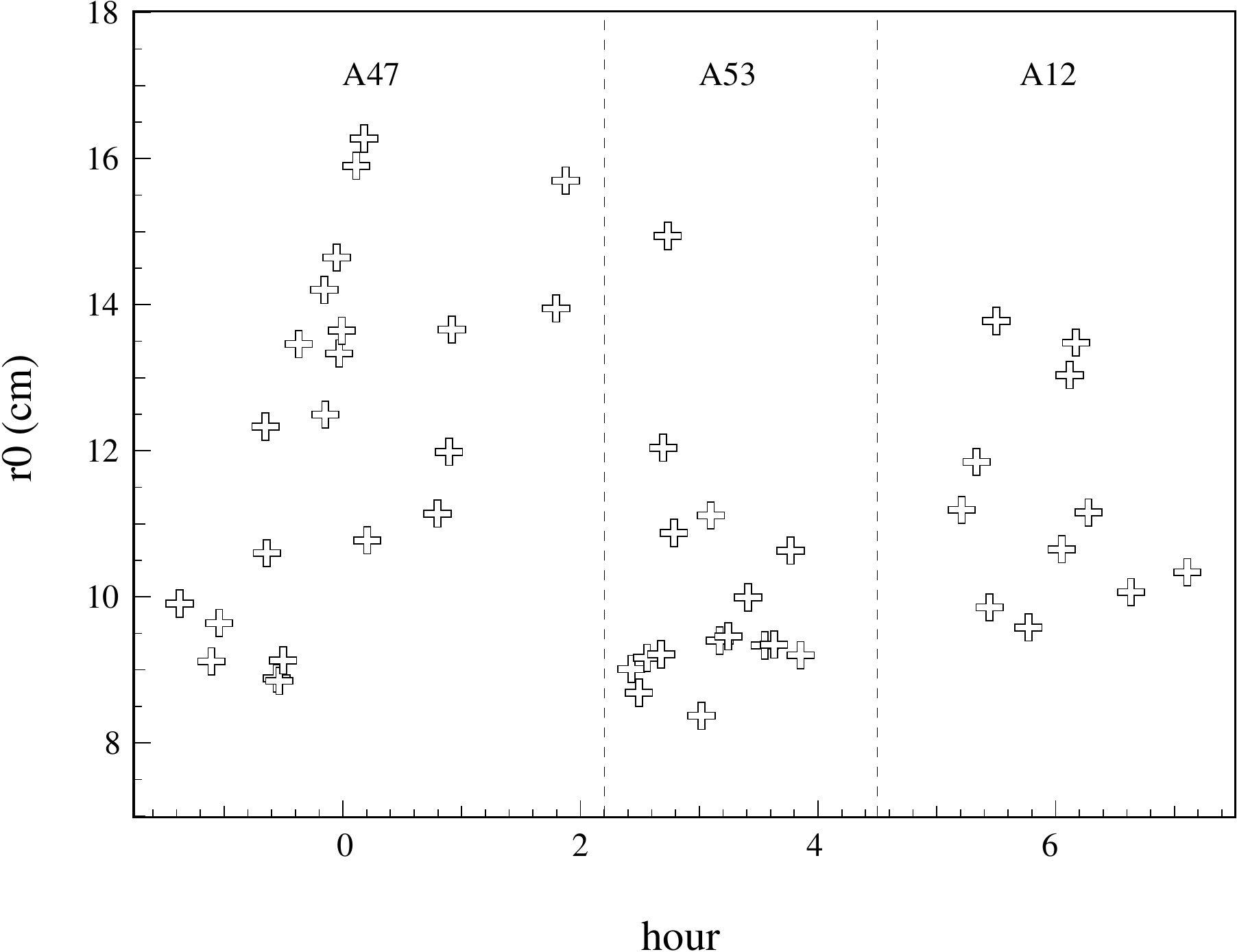}
\caption{$r_0$(cm) at 500~nm from WFS data versus local time (hour) during the night of the 27th September 2010. Dashed lines separate asterisms.}
\label{r0VStime}
\end{figure}

Figure~\ref{r0VStime} plots the $r_0$ estimated from off-axis WFS data (unbiased from noise and aliasing, see Sect.~\ref{seeing}) versus the local time (in hours, negative before midnight). The $r_0$ value is not rescaled to zenith.  Dashed lines represent the asterism change during the night. Asterism A47 was observed from the beginning of the night to \hm{02}{00} local time, A53 from \hm{02}{00} to \hm{05}{00} and A12 from \hm{05}{00} to the end of the night. The median value is $r_0=11.33$~cm giving a median seeing of $0.91\arcsec$ at $500$ nm. Worst and best seeing were respectively $1.23\arcsec$ and $0.63\arcsec$.

\subsection{IR camera images}
\begin{figure}
\centering
\includegraphics[width=9cm]{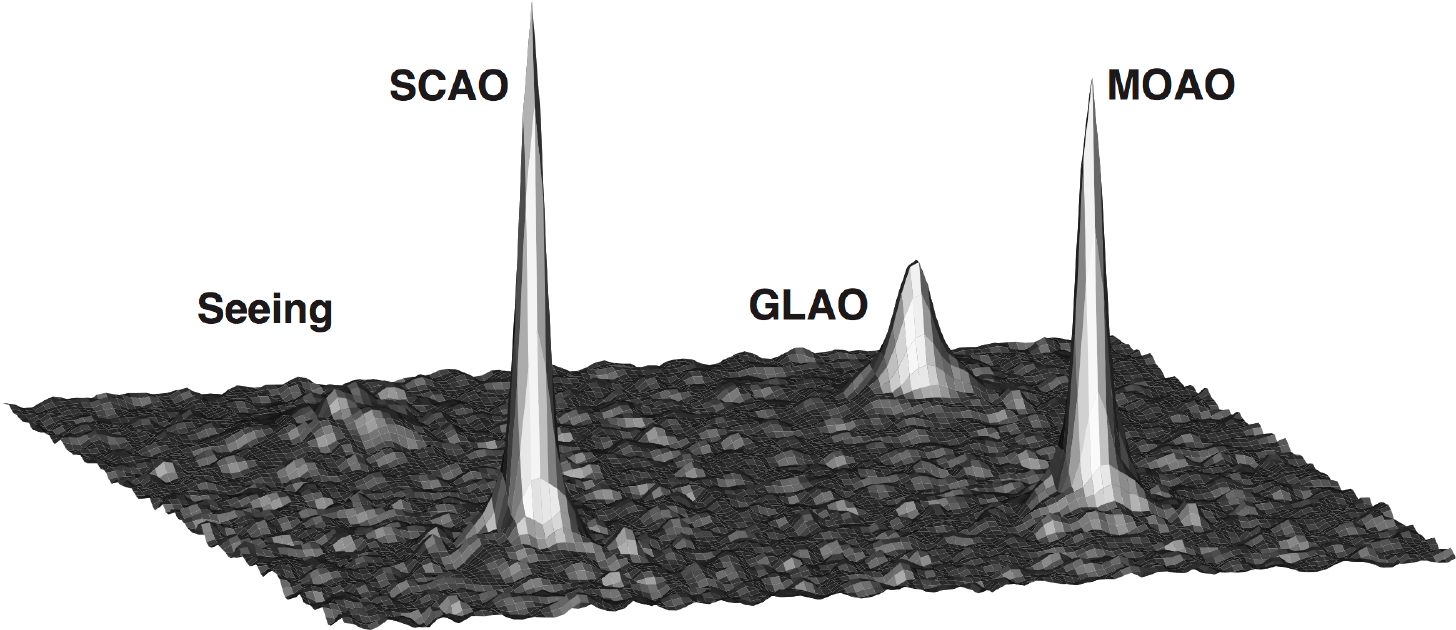}
\caption{IR image comparison at $\lambda=1\,530$~nm. The four images of 30 seconds exposure each were taken at \hms{00}{59}{18} (Seeing), \hms{00}{42}{10} (GLAO), \hms{00}{29}{22} (MOAO) and \hms{00}{32}{28} (SCAO). Measured SR are respectively: 1\%, 9\%, 19.4\% and 23.8\%.}
\label{SR_comparison3D}
\end{figure}

Figure~\ref{SR_comparison3D} shows an example of IR image comparison at $\lambda=1\,530$~nm for the three AO modes tested on CANARY plus a seeing limited image. The four images of 30 second exposure each were taken respectively at \hms{00}{59}{18} (Seeing), \hms{00}{42}{10} (GLAO), \hms{00}{29}{22} (MOAO) and \hms{00}{32}{28} (SCAO). Measured SR are respectively, 1\%, 5.3\%, 19.4\% and 23.8\%. As already underlined by \citep{Gendron11}, the MOAO performance is close to the SCAO one. In particular, despite the strong ground layer observed during the night, we see that MOAO performs much better than GLAO. The following subsections will give more insights into this result.

%

\subsection{Turbulence profile}
\label{profile}
During our observations, the $C_n^2(h)$ retrieval step (i.e. the \emph{Learn} phase of the L\&A algorithm), was limited for practical software reasons (computation time) to only three fitted layers. Since September 2010, we have significantly improved the computation speed of the \emph{Learn} step \citep[see][]{martinSpie2012}. On-sky, we were jointly fitting, at the same time, the strengths and altitudes of 3 layers and the 4 WFS on-sky positions (observing directions). With the new fitting procedure we now employ a series of altitudes $h_i$, regularly spaced by $\Delta_h$, ranging from ground to $h_{max}$
(see Sect.~\ref{tomoreso}) and fit only their strength value $C_n^2(h_i)$. The positions of the WFS are measured by the Target Acquisition System encoders in the focal plane of the telescope. The fit is now more robust and thanks to an increase in speed, we are now able to fit up to 15-20 layers in a few tens of seconds.

\begin{figure}
\centering
\includegraphics[width=8cm]{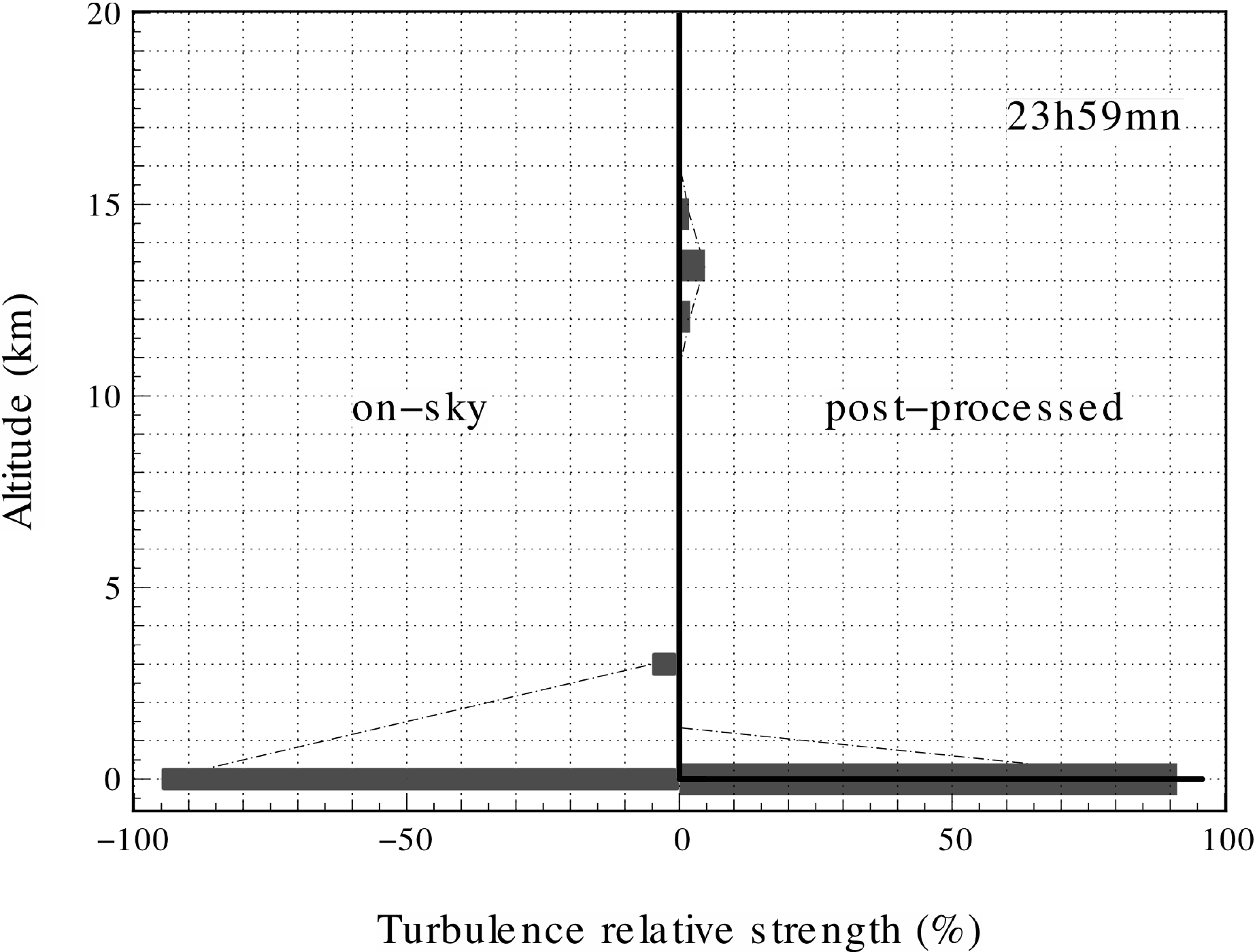}
\caption{$C_n^2(h)$ retrieved on-sky at \hm{23}{59} on Asterism $\#$47 (left). More detailed $C_n^2(h)$ with 15 layers (right) obtained after post-processing of the same data set.}
\label{23h59}
\end{figure}

\begin{figure}
\centering
\includegraphics[width=8cm]{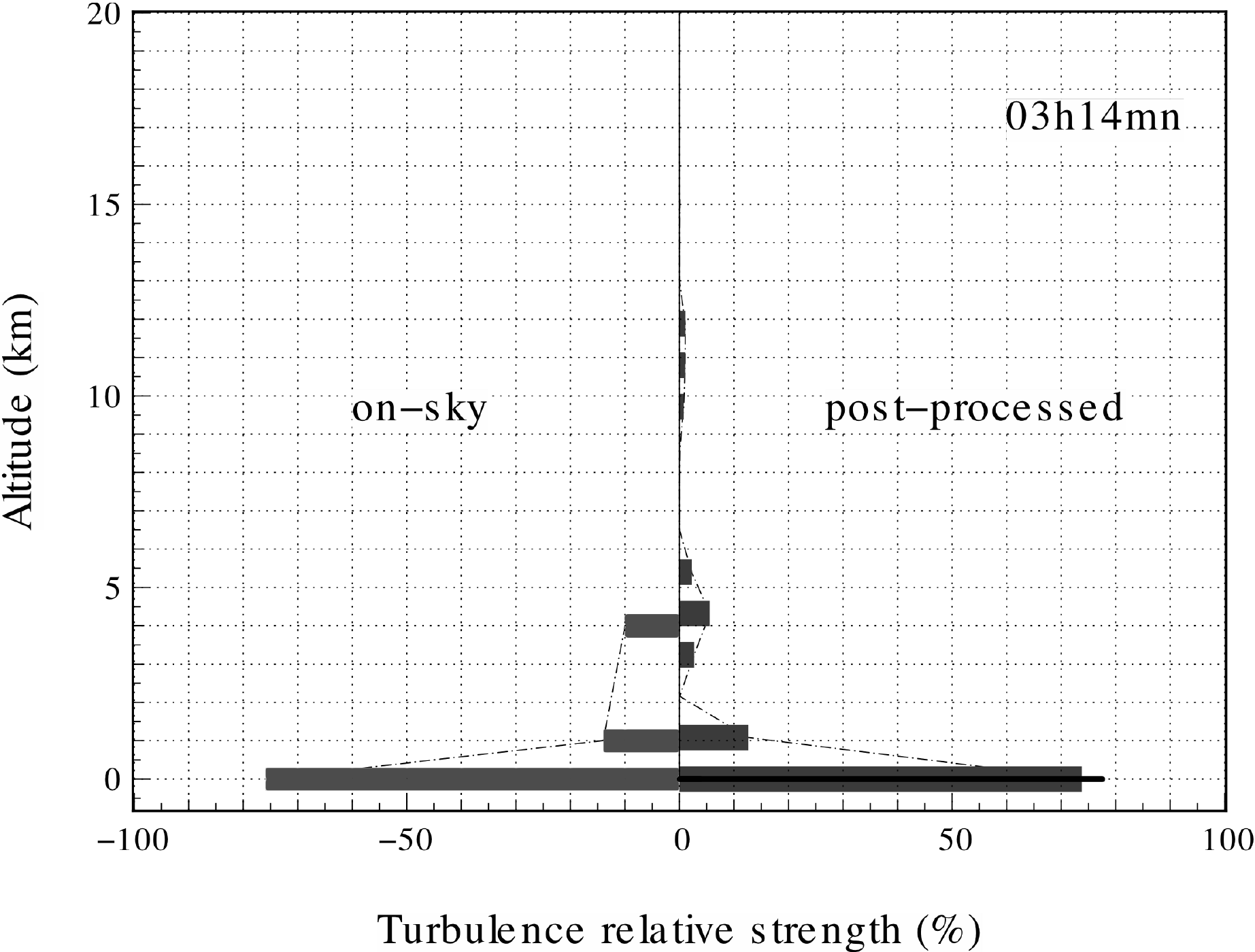}
\caption{$C_n^2(h)$ retrieved at \hm{3}{14} on Asterism $\#$53 (left: on-sky; right: post-processed).}
\label{03h14}
\end{figure}

\begin{figure}
\centering
\includegraphics[width=8cm]{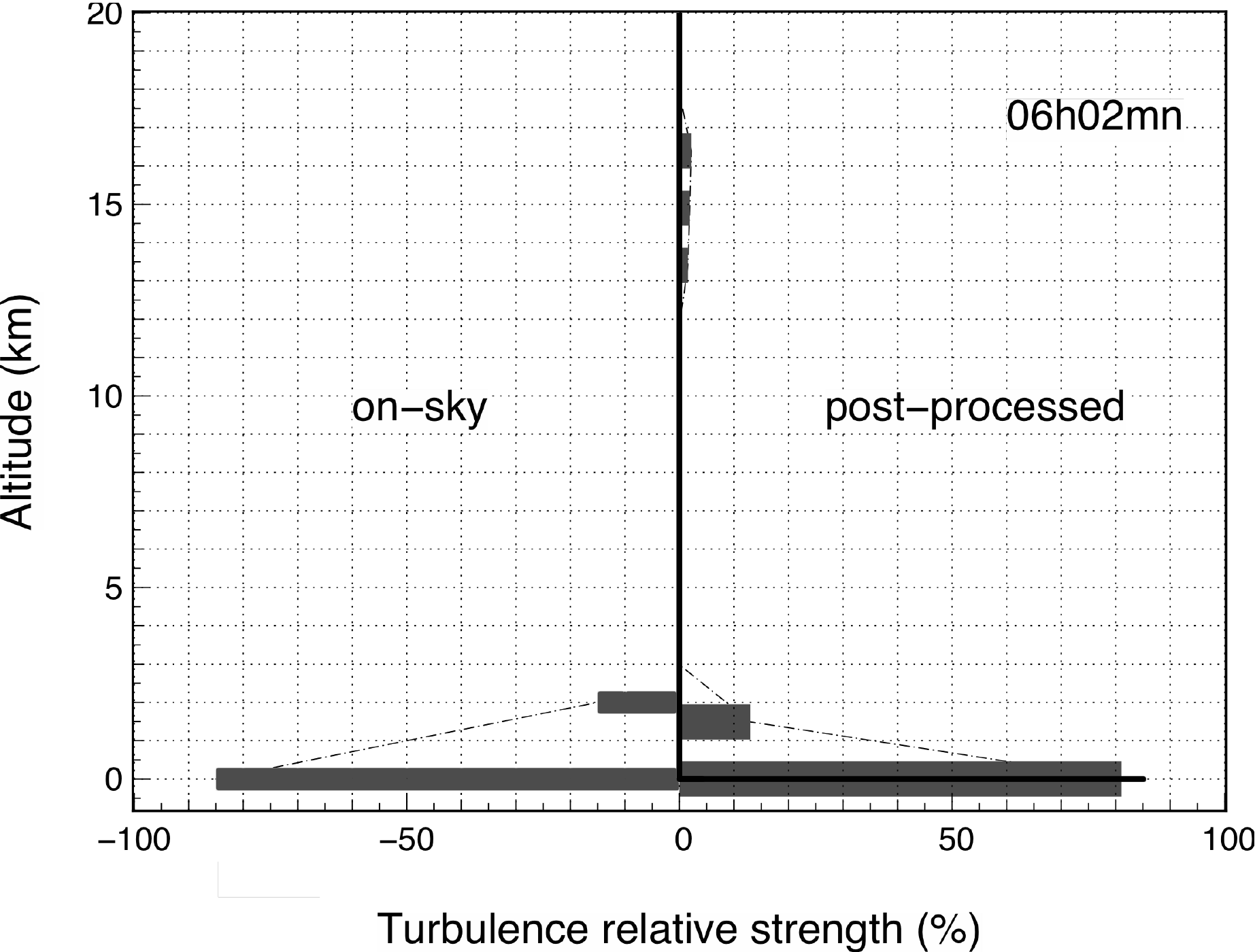}
\caption{$C_n^2(h)$ retrieved at \hm{6}{02} on Asterism $\#$12 (left: on-sky; right: post-processed).}
\label{06h02}
\end{figure}

Post-processing the Phase A on-sky data allows us to retrieve detailed $C_n^2(h)$ profiles with up to 15 layers measured between 0 and 20~km (depending on the asterism geometry). We have selected one \emph{disengaged slopes} data set per asterism to illustrate the identification of the turbulence profile and the estimation of the error budget. Table~\ref{TomoSky} summarises the parameters of these synchronised data sets taken at \hm{23}{59} (using 28\,000 frames, acquisition time of \hm{3}{06}), \hm{3}{14} (58\,000 frames) and \hm{6}{02} (28\,000 frames) on asterisms A\#47, A\#53 and A\#12, respectively.
Figures~\ref{23h59}, \ref{03h14} and~\ref{06h02} present the $C_n^2(h)$ measured during the on-sky observations (left) and post-processed (right) using these sets of \emph{disengaged slopes}. The turbulence profiles have not been rescaled by the airmass to maintain the profile as it was observed in the line of sight of CANARY.

At \hm{23}{59} (Fig.~\ref{23h59}), we measured more than 95\% of the turbulence below 1\,340~m (best resolution achievable on asterism A\#47, see also Table~\ref{TomoResolution}) during the on-sky operations. A weak layer of 5\% was also detected at $\approx$ 3\,000~m. After post processing the data with a higher vertical resolution sampling, we measured the ground layer contribution at 91\% and one high layer at $\approx$13\,500~m of 9\% (right). 
On the \hm{3}{14} dataset (Fig.~\ref{03h14}), we measured during the observations a ground layer contribution at 77\% (74\% post processed), a 14\% layer contribution at 1\,080~m (11\% post processed) and 1 layer at 10\% at $\approx$4\,000~m (spread into 3 layers of 10\% contribution with the new 15 layer profile). We additionally measured with the post processing 3 very weak layers around 11km (few percent) which was not measured and taken into account on-sky.
At \hm{6}{02} (Fig.~\ref{06h02}), we measured on-sky 85\% in the ground layer contribution (82\% post processed) and one layer at $\approx$2\,000~m of 15\% (13\% at 1\,490~m + 3~small layers at $\approx 15$~km of a total contribution of 5\% with post-processing).
We can see on these plots that even with only three fitted layers we were able to account for most of the turbulence distribution in our computed estimators. But it is also clear that we have a significant model error in our results because of the missing high altitude layers in the estimators. We will quantify this error in Sect.~\ref{details}.

%
%
%
%
%

\subsection{Tomographic error}
\begin{figure}
\centering
\includegraphics[width=8cm]{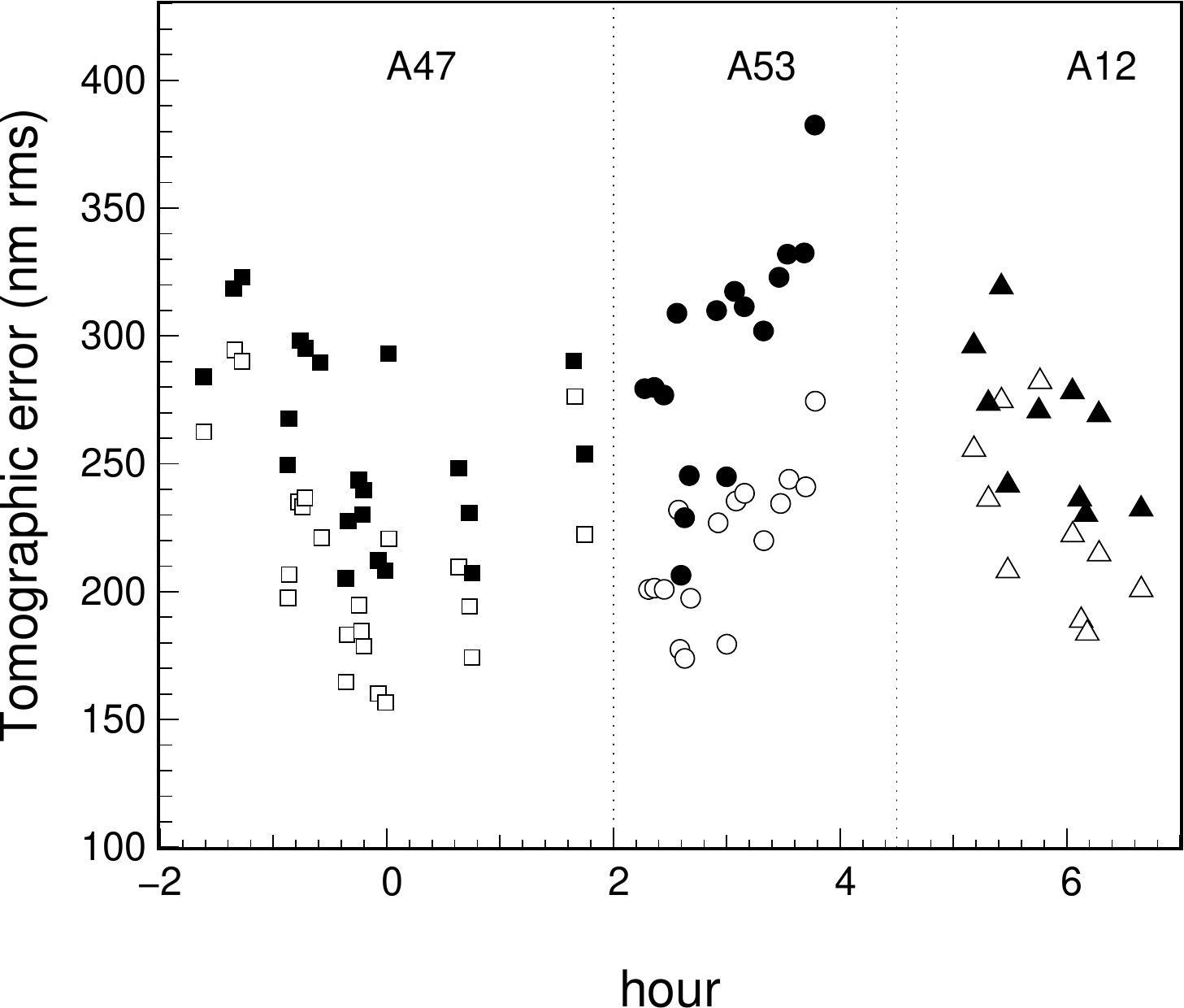}
\caption{Tomographic error (nm rms) measured using \emph{disengaged slopes} versus local time (hour). Asterisms A47, A53 and A12 are respectively represented with crosses, circle and triangles. For each asterism we plot the MOAO and GLAO reconstruction respectively represented in plain and blank points.}
\label{tomoVStime}
\end{figure}

Figure~\ref{tomoVStime} shows the tomographic error  versus the local time. We used here sets of \emph{disengaged slopes} to compute the tomographic error (see Eq.~\ref{tomoeq}). The graph plots the tomographic error for GLAO cases (crosses) and MOAO cases (L\&A with 3 reconstructed layers: circles).
At any time, the L\&A reconstruction performs better than the GLAO reconstructor. The median tomographic error for all data taken over the night gives $\sigma_{tomoMOAO}=$216~nm rms for MOAO while the GLAO reconstruction gives $\sigma_{tomoGLAO}=$270~nm rms. On average throughout the night, MOAO reconstruction performed 160~nm rms better than the GLAO case. Most of the fluctuations in the tomographic error observed in Fig.~\ref{tomoVStime} are due to the large fluctuations in the seeing conditions during the night.

\begin{figure}
\centering
\includegraphics[width=8cm]{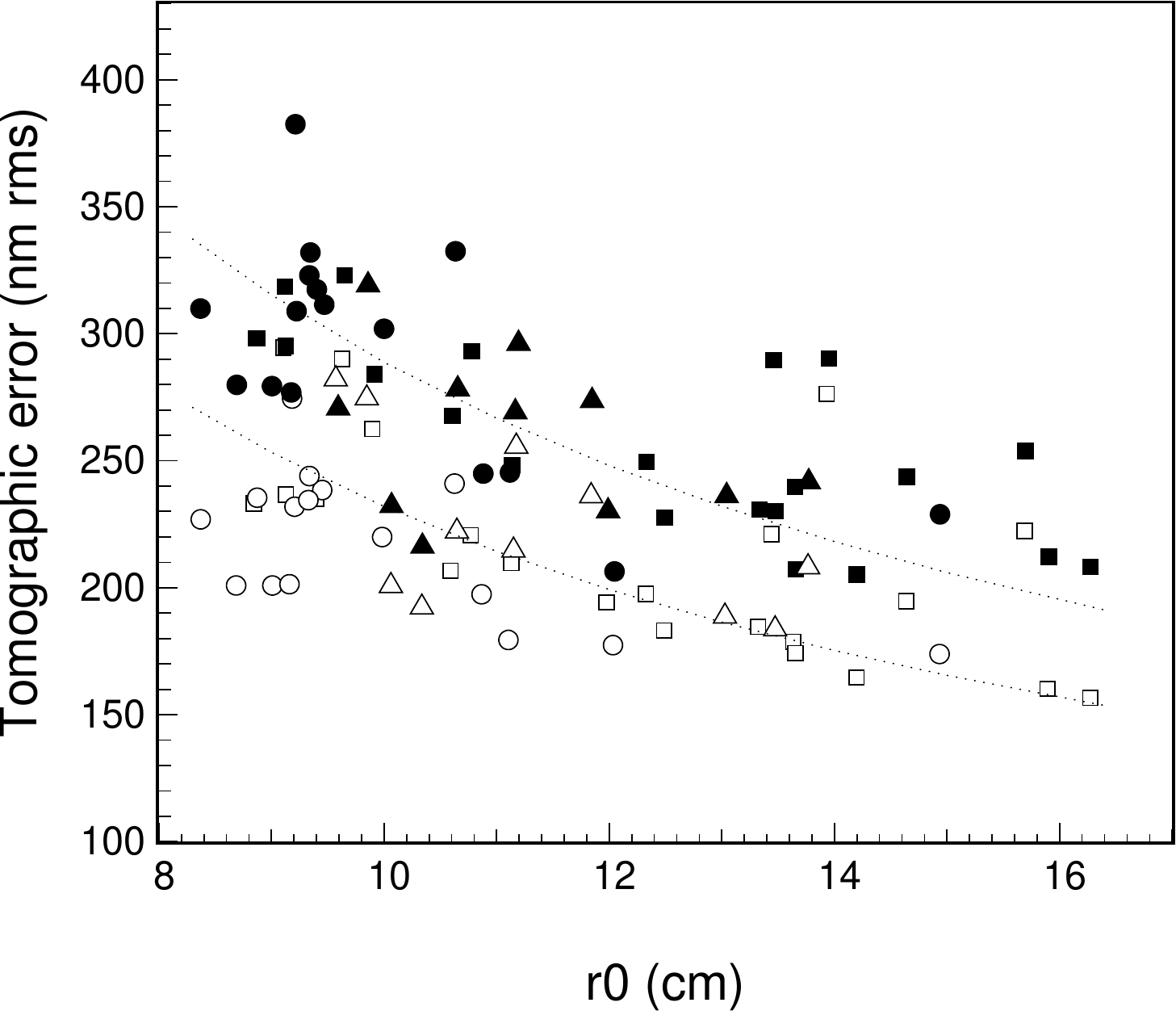}
\caption{Tomographic error measured with MOAO (circles) and GLAO (crosses) estimator versus r$_0$(cm). Dashed lines represent a fit to each dataset by a law $\propto \left(D/r_0\right) ^{5/6.}$}
\label{tomoVSr0}
\end{figure}

Figure~\ref{tomoVSr0} plots the tomographic error as function of $r_0$ value. We superimpose the expected $\left(D/r_0\right)^{5/6}$ law showing the overall behavior of this error term. This trend is globally verified by our estimates of this error. One can see the improvement of  the L\&A tomographic reconstruction versus the GLAO mode. The large scatter of the MOAO results is due to the large number of configurations tested during the night.
Many MOAO datasets were taken when system parameters were not optimal and the observing conditions were also evolving. 

\begin{figure}
\centering
\includegraphics[width=8cm]{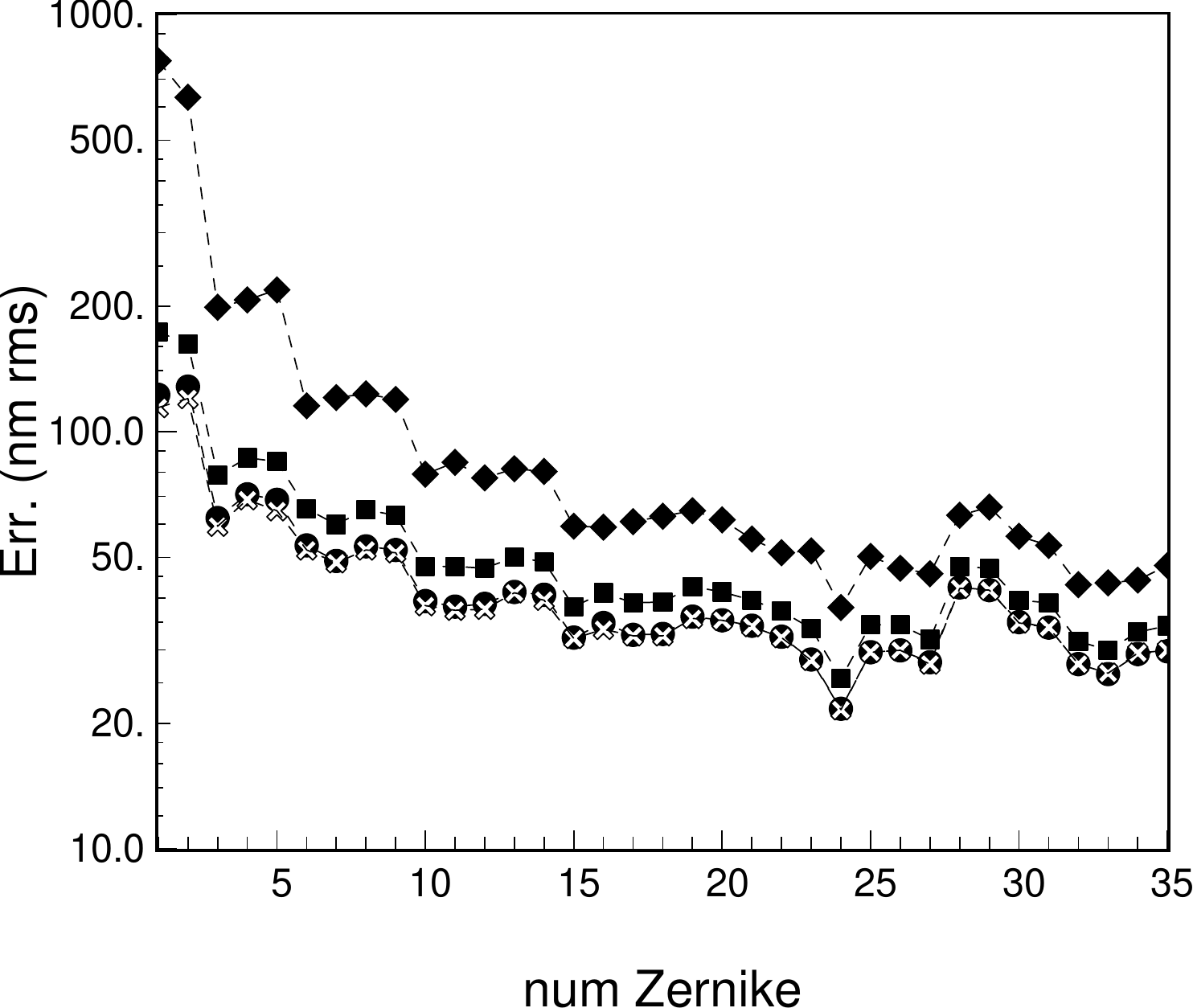}
\caption{Zernike expansion coefficient root mean square (log scale) of uncorrected wavefront (black/diamond), tomographic reconstruction with GLAO (black/square) MOAO correction with 3 reconstructed layers (black/circle) and 15 reconstructed layers (cross), for the \emph{disengaged} slopes dataset of \hm{3}{14}.}
\label{03h14zer}
\end{figure}

\begin{figure}
\centering
\includegraphics[width=8cm]{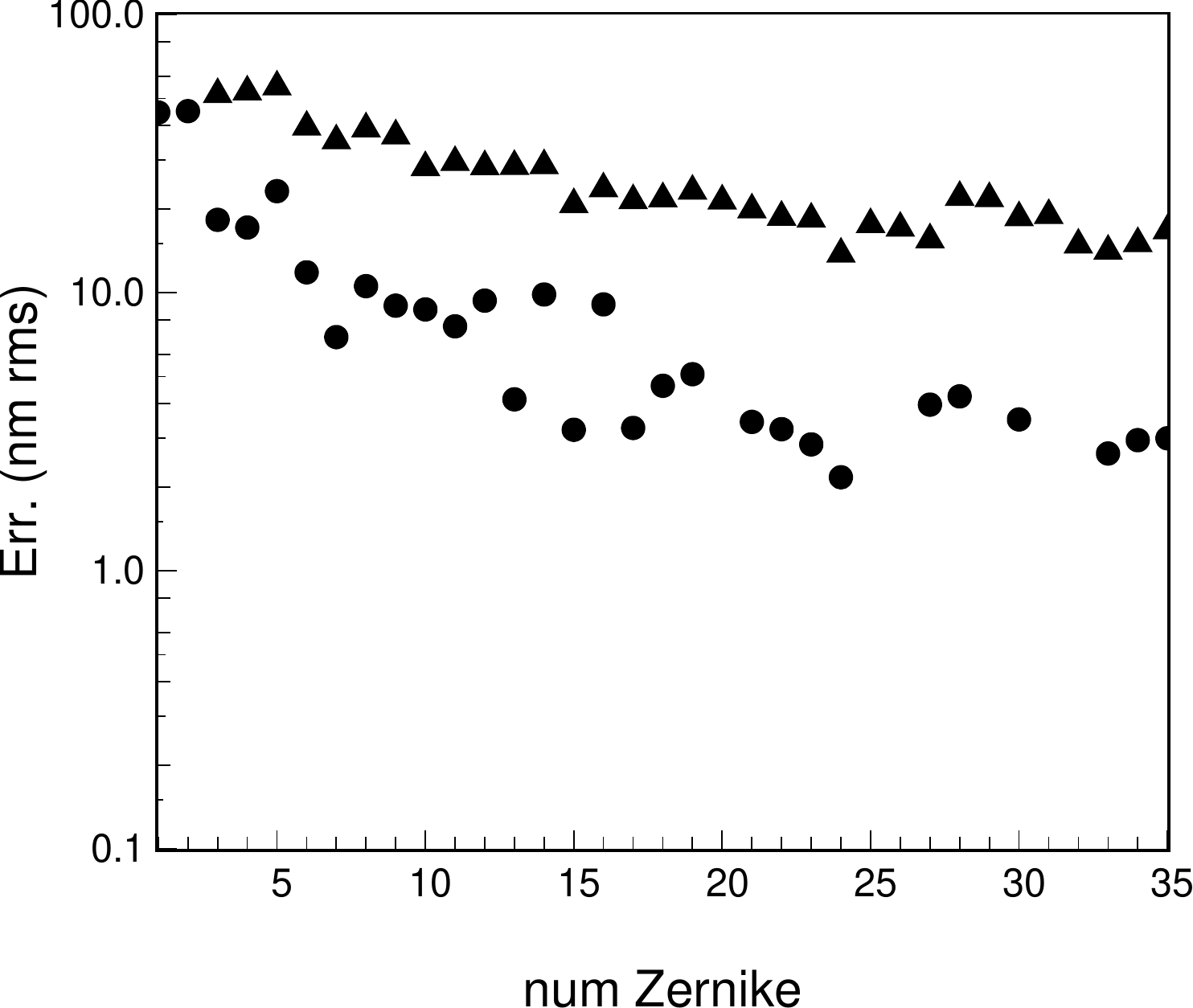}
\caption{We plot the quadratic difference between the GLAO and 3 layer MOAO reconstruction (triangles) as a function of the Zernike number using the same dataset as Fig \ref{03h14zer}. The 3 layer reconstruction improves the tomographic reconstruction by 215nm rms in total. The quadratic difference between the 3 layer and 15 layer MOAO reconstruction is illustrated with circles. The 15 layers reconstruction allows us to reduce the tomographic error by 84nm rms.}
\label{03h14zerDiff}
\end{figure}

Figures \ref{03h14zer} represent the statistics of Zernike expansion coefficients on the data set of \emph{disengaged} slopes at 03h14. We compute the Zernike expansion of the uncorrected turbulence (black/diamonds), the GLAO reconstruction (black/square) MOAO correction with 3 reconstructed layers (black/circle) and 15 reconstructed layers (empty/cross),
The uncorrected wavefront error is measured at a level of 1\,139~nm rms. As expected, we recognise the Zernike expansion sorted by radial orders (order 1 with tip and tilt, order 2 containing the defocus and the two astigmatisms, etc). The GLAO reconstruction gives a tomographic error of 314~nm rms while the optimized MOAO reconstruction with 3 layers gives 229~nm rms (with the MOAO estimator used on-sky). We also use the optimized 15 layers $C_n^2(h)$ measured a posteriori (see Fig.~\ref{03h14}) to compute a 15 layer tomographic estimator, apply it to the off-axis data and compare the prediction to the real measurements made on-axis by the TS. The post-processed estimator increases performance by reducing the model error, resulting in a wavefront error of 213~nm rms. 
Figure ~\ref{03h14zerDiff} illustrates the quadratic difference between the GLAO and MOAO reconstruction with 3 layers (black/triangles) as function of the Zernike number. One clearly sees the improvement for all modes using the MOAO reconstructor with 3 layers (tomographic error reduced by 215nm rms).
Similarly Fig. ~\ref{03h14zerDiff} illustrates the quadratic difference between the MOAO reconstruction with 3 layers (black/circles) and MOAO reconstruction with 15 layers (black/triangles). For this dataset the total improvement by reconstructing 15 layers instead of 3 could be reduced by 84nm rms.

%


Table~\ref{TomoSky} summarizes all the numbers for 3 data sets of \emph{disengaged} slopes respectively at 23h59, 3h14, and 6h02 and highlights the quadratic differences between error terms.
It shows that the 15 layer profile  increases the performance of the tomographic reconstruction compared to the 3 layer reconstruction  by an amount ranging between 80 and 60~nm rms. 
This comparison reveals the impact of the model error in the tomography error term, even in our observing conditions with a dominant ground layer. 
Correcting with the 3 layer estimator brings an improvement of 164, 215 and 173~nm~rms over the GLAO estimator for the three data sets. 

\begin{table}
\caption{Three disengaged data sets for the three asterisms and the corresponding tomographic error computed with a ground layer, 3 or 15 layer tomographic estimator (in nm rms).}
\label{TomoSky}
\centering   
\begin{tabular}{l c c c}
	\hline
	\hline
Asterism	      &       $\#$A47 & $\#$A53 & $\#$A12 \\
	  \hline    
Disengaged slopes hour	   &        \hm{23}{59} &  \hm{3}{14} & \hm{6}{02}  \\
Recorded frame number  & 28\,000 & 58\,000 & 28\,000 \\
$r_0$ (cm)  & 14.1  & 9.7  &  10.6 \\
Total turbulent rms error   &    1052          &          1139      &   1133     \\
$\sigma_{Tomo+Model}$ (GLAO)    &      249          &            314      &    286      \\
		$\sigma_{Tomo+Model}$ (3  layers)    &      187          &          229        &    228     \\
 $\sigma_{Tomo+Model}$ (15 layers)    &      175          &            213      &    219      \\
quadratic diff. (3-15 layers)    &      66          &            84      &    63      \\
quadratic diff. (GLAO-3 layers)    &      164     &         215      &     173      \\
	\hline
\end{tabular}
\end{table}

%

\subsection{Open loop error}
\begin{figure}
\centering
\includegraphics[width=8cm]{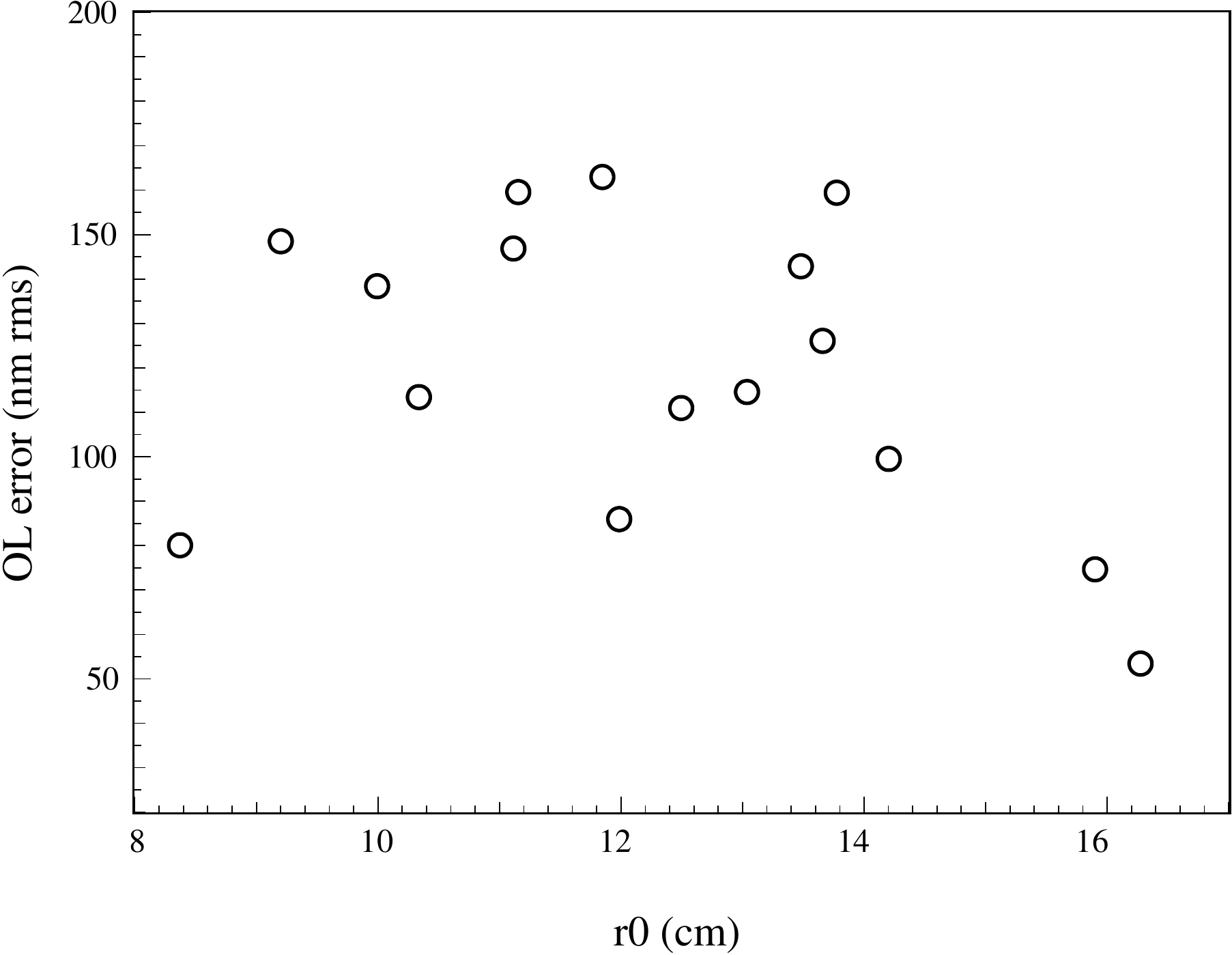}
\caption{Openloop error versus $r_0$.}
\label{OLVSr0}
\end{figure}

Open loop error is represented as a function of $r_0$ in Fig.~\ref{OLVSr0}.
Open loop error estimation requires \emph{engaged slopes} to be processed with \emph{disengaged slopes} taken together within a short time scale. Only 16 sets of data are available with engaged and disengaged slopes close enough (in time) to compute a valuable error budget during the night. We estimate that open loop error ranged between 160 and 50~nm rms during the night. Because of the method used to estimate this error (computed from the quadratic difference of numerous estimated error terms), we cannot reliably draw conclusions about the behavior with respect to $r_0$.

The open-loop error (or go-to error) was measured on the bench as typically lower than 5\% of the total corrected wavefront \citep{kellererDM2012}. For the three data sets of Table \ref{TomoSky} this gives an error lower than 35nm rms.

\subsection{Field static aberration error}
\begin{figure}
\centering
\includegraphics[width=8cm]{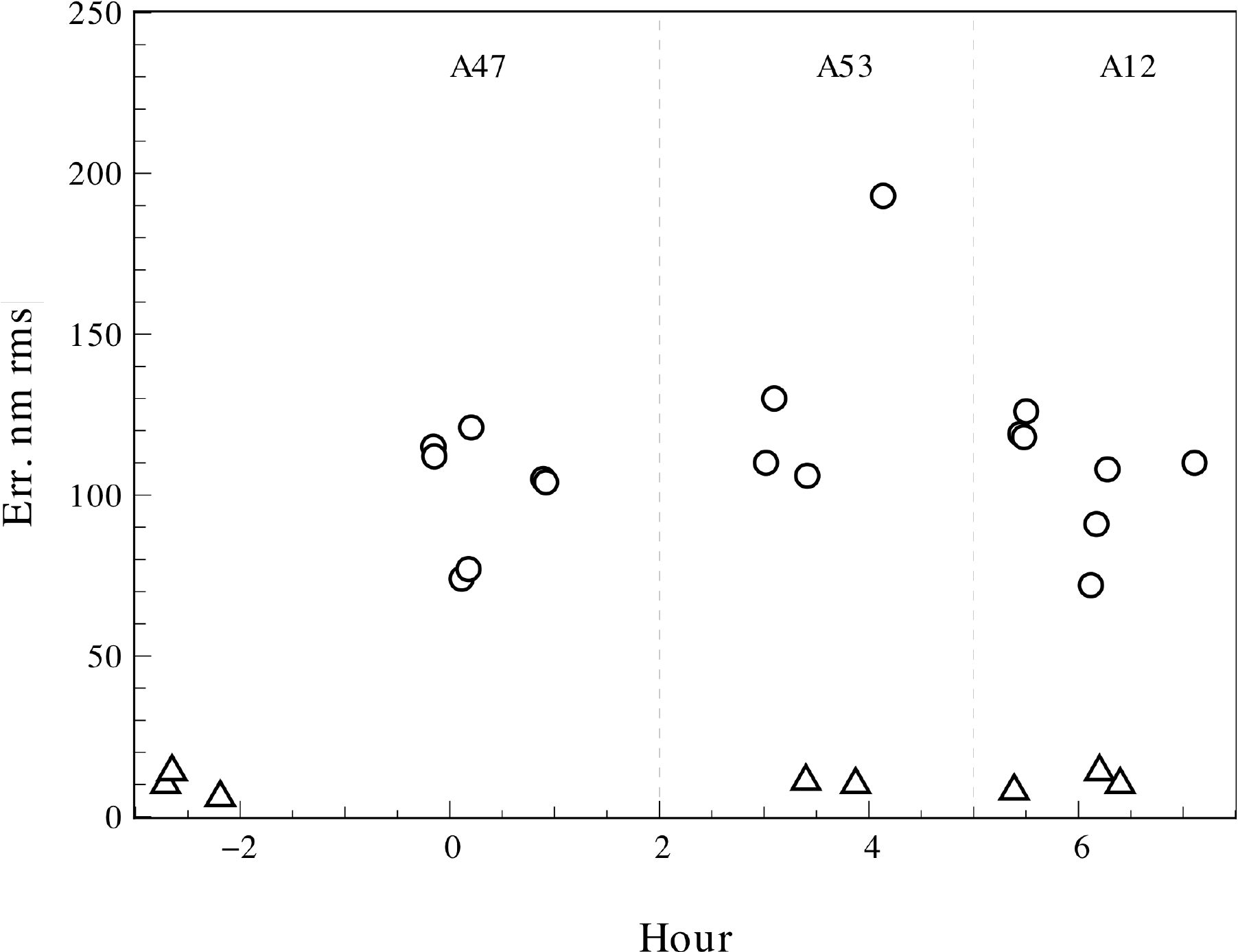}
\caption{Wavefront error  (in nm~rms) due to static aberrations when measured in SCAO (triangles) and MOAO (circles) configurations versus local time. }
\label{staticVStime}
\end{figure}

The measured field (quasi-static) aberrations versus the local time is presented in Fig.~\ref{staticVStime}. Residual static aberrations are very small $\approx $10~nm rms in the SCAO case (triangles). This value can be explained by the larger number of freedom measured by the TS than the number really controlled by the DM. The measured field aberrations in MOAO (circles) is 110~nm rms averaged over all the night. This value is quite large compared to other error terms and should not be underestimated while designing an MOAO instrument. Within this value, we estimated the order of magnitude of the contribution of the DM creep to be around 60nm rms. It also shows that MOAO requires a regular update of these field aberrations either by calibrating them on-sky (CANARY scheme) or by computing them using a model of the telescope aberrations.

\section{Error budget comparison}
\label{details}
In this section we present the error budget computed on-sky at 3 different times during the night,  each made on a different asterism in order to characterize the performance on 3 different geometrical and atmospherical conditions.
Each case requires both a disengaged and a sequential engaged data set, in order to be able to perform the full error budget. The 3 disengaged sets are the ones presented in Table~\ref{TomoSky} and described in the previous section. The hours for the different data sets are:

\begin{tabular}{ccc}
\hline
Asterism  &  Disengaged &   Engaged \\
\hline
A\#47        &   \hm{23}{59}  &    \hm{0}{10} \\
A\#53       &    \hm{3}{14}  &     \hm{3}{24}  \\
A\#12       &    \hm{6}{02}  &     \hm{6}{07}  \\
\hline
\end{tabular}

In order to compare to the on-sky results, we also perform three numerical simulations for the same conditions and we compute the corresponding error budget and IR images. We present the simulation parameters in the next section.

\subsection{Simulation parameters}
We used the end to end adaptive optics simulation program called {\em Yorick Adaptive Optics} (YAO)  written by \citep{YAO} to produce series of \emph{engaged} and \emph{disengaged} data sets of slopes for each of the 4 WFS (including TS). These data sets mimic the data that were acquired on-sky with CANARY.
The YAO code was modified to perform a simulated MOAO correction, in particular using the tomographic estimator $M_{ct}$ (see Sect.~\ref{implementation}) and the open loop integrator scheme as described in Sect.~\ref{Control}. Finally, the 2 sets of data (\emph{engaged} and \emph{disengaged} slopes) produced by YAO are used as inputs to our on-sky data reduction software.
We emphasize that we used the same data pipeline software to compute both the on-sky error budget and the simulated error budget in order to avoid any bias in the data reduction.

YAO is able to simulate a multi-layer turbulence profile by generating independent phase screens for each layer. 
We have chosen the 15-layer profile deduced from the post-processed on-sky data (see Fig.~\ref{23h59}, \ref{03h14} and \ref{06h02}) as the simulated $C_n^2(h)$. Simulated WFS positions in the field of view are the ones deduced from the measurement made during the on-sky observations by WFS position encoders within the Target Acquisition System.

We simulate the on-sky WFS configuration with $7\times7$ sub-apertures and $16\times16$ pixels per sub-aperture. The centroids of spots are computed with the same centroiding method as that used on-sky, i.e. selecting the 10 brightest pixels in a sub-aperture and zeroing the others.
The magnitudes of the guide stars are adjusted to match with the WFS noise variance measured directly from the on-sky data. The variance of the noise on the simulation dataset is therefore identical to the real noise encountered on-sky, with both measured as described in~\ref{Noise}.

The sampling frequency is simulated at 150~Hz. In order to simulate a loop latency of 1.5 frames, we ran two numerical simulations to compute the bandwidth error at 1 and 2 frames delay (YAO only simulates latencies of an integer number of frames) and linearly interpolated the bandwidth error at 1.5 frames. We simulated 7\,000 iterations ($\approx$~45~seconds at 150~Hz) to produce typical data sets on all the WFS. 
The estimation of the average wind speed  was deduced from the on-sky datasets by measuring the FWHM of the temporal autocorrelation of disengaged slopes data, and we reproduced the same in the simulation by tuning the speed. The simulated wind speeds were set respectively to 6, 6 and 8~m~s$^{-1}$ as measured from the datasets at \hm{0}{10}, \hm{3}{24} and \hm{6}{07}. 
The loop gain is respectively set to 0.8, 0.6 and 0.6 corresponding to the values used in these 3 sets. The wind speed is kept the same for all the layers in the simulation. 
Other parameters of the simulation are identical to the CANARY configuration used on-sky (see Sect.~\ref{instrument}) and they are summarized in Table~\ref{Simusparams}. 

\begin{table*}
\caption{Simulations parameters}  
\label{Simusparams}
\centering   
\begin{tabular}{lccc}
	\hline
	\hline
 Asterism &   $\#$A47 & $\#$A53&   $\#$A12 \\
         \hline
         frequency  (Hz) &   150     &    150     & 150 \\
       	 delay (frames) &   1.5     &    1.5     & 1.5 \\
	 simulated iterations&   7\,000     &    7\,000     & 7\,000 \\
         total  $r_0$ (cm)   &   16.3     &    10.0     & 13.0 \\
         $C_n^2(h)$   &   Fig.~\ref{23h59} right    &    Fig.~\ref{03h14}   right  & Fig.~\ref{06h02} right\\

         windspeed (m/s)   &   6     &    6    & 8 \\
         loop gain   &   0.8     &    0.6     & 0.8 \\
         nb pix/sub ap.   &   $16\times16$     &    $16\times16$     & $16\times16$ \\
             nb Brightest pix/sub ap.   &   10     &    10     & 10 \\
              WFS pixsize (\arcsec/pix)  &   0.26     &    0.26    & 0.26 \\
              star magnitude (1,2,3,TS)  &   [10.2,11.7,11.4,12.5]     &     [11.2,12.3,13.0,12.8]    &   [13.2,13.8,11.9,10.8]  \\
	    WFS noise (nm rms) &   [70,127,108, 162]     &     [99,142,187,176]    &   [206,328,137,85]  \\
	    RON (e$^-$) &   0.5     &    0.5    &   0.5  \\
	 \hline
\end{tabular}
\end{table*}

\subsection{On-sky error budget}
In this section we discuss the on-sky error budget computed from \emph{engaged slopes} listed at the beginning of Sect.~\ref{details}.
We recall that the \emph{engaged slopes}  are used to compute, $r_0$, $\sigma_{ErrTS}$, $\sigma_{tomo\ Noise}$, $\sigma_{AliasAlt}$, $\sigma_{Alias_{Ground}}$, $\sigma_{AliasAltTS}$, $\sigma_{BW}$, $\sigma_{noiseTS}$, $\sigma_{Fit}$, and  $\sigma_{Field\_stat}$.
The important term $\sigma_{Tomo}$ is missing to this list: we need sets of \emph{disengaged slopes} recorded a few minutes before the engaged ones to determine it (using Eq.~\ref{tomoeq}) and we then rescale it according to the $r_0$ value found on the disengaged set.
Then we use Eq.~\ref{ErrOL} to estimate the open loop term $\sigma_{OL}$. 
Finally we estimate the overall error budget $\sigma_{ErrIR}$ on the IR camera using Eq.~\ref{IRTotal}. IR images (30~s exposure) were also recorded at a time very close to the engaged slopes.
Results are presented in columns labelled \emph{on-sky} in Table~\ref{err} and split in three parts, each for their respective couple of data sets of engaged and disengaged slopes.

\begin{itemize}
\item 
The tomographic error (including the turbulence model error) is estimated on-sky at $\sigma_{Tomo}$ = 156, 219 and 188~nm rms respectively on asterisms A47, A53 and A12. Those 3 values are those already given in Table~\ref{TomoSky}, but now rescaled with the $r_0^{-5/3}$ of the engaged sets.
We recall they were computed using the 3-layer tomographic estimator used on-sky (Figs.~\ref{23h59}, \ref{03h14}, and \ref{06h02} on the left for the corresponding profiles).
\item 
The estimated open loop error ($\sigma_{OL}$) ranges between 55 and 140~nm rms. This error term also includes all the other error terms not identified in the error budget. 
\item 
Noise propagated through the estimator ($\sigma_{TomoNoiseFilt}$, Eq.~\ref{TomoNoisefilt}) ranges between 48 and 97~nm rms. 
In spite of the large noise terms measured on the individual WFSs (up to 328~nm rms, see Table~\ref{Simusparams}), the noise actually injected in the loop is still one of the weakest error terms.
This is explained firstly because we take advantage of  multiple wavefront sensing directions allowing the averaging of the noise on the 3 off-axis directions. Secondly, the tomographic estimator takes into account the average noise variance in each sub-aperture (diagonal of the covariance matrix $C_{OffOff}$) and consequently deals optimally with a noisy WFS. Finally we use a temporal controller to filter part of the noise out.
\item 
We estimate the aliasing contribution at the ground ($\sigma_{AliasGround} $) at 83 to 132~nm rms, and in altitude ($\sigma_{AliasAlt}$) at 15 to 28~nm rms. Because of the predominance of the turbulence at ground level in the $C_n^2$ profile (never less than 75\%), the ground layer contribution to the aliasing dominates. 
\item 
The measured on-sky bandwidth error lies from 115 to 142~nm rms. The bandwidth error is the highest in A53 in particular because the loop gain was set only to 0.6 instead of 0.8 for the other cases, and the seeing was the worst.
\item 
We estimate the fitting error at 138, 206 and 165~nm rms which makes it the second largest contributor to the  error on CANARY after the tomography.
\item 
We observe a non negligible contribution (between 77 and 106~nm rms) of the field static aberrations to the error budget even when using our dedicated calibration procedure to extract such a term from the measurements. This contribution may come from the DM creeping effect and from the evolution of the field aberrations during the observations through field derotation or telescope flexures.
\item 
Best SR errors were measured on the bench with the calibration sources at 115~nm rms and were considered as fixed during the night of observation (see Sect.~\ref{NCPA}). 
\end{itemize}
Finally, the estimated total error budget for the IR camera ($\sigma_{ErrIR} $) gives 297, 419 and 357~nm rms. Converting these to Strehl ratio using the approximate formula 
\begin{equation}
\mathrm{SR} \approx \exp(-(2\pi\sigma_{ErrIR}/ \lambda)^2)
\end{equation}
at $\lambda = 1\,530$~nm, we find the estimated SR: 22.6\%, 5.2\% and 11.7\%. 
Real IR images were recorded a few tens of seconds from the dataset of slopes. We measure on these images SR of 20.1\%, 10.3\% and 16.4\% ($\pm$2\%) respectively. The observed discrepancy is mainly due to the pessimistic estimation of the SR using the exponential formula from the error budget. This formula gives much better estimation when SR is higher, typically larger than 30\%. We will come back to this point after the presentation of the simulated error budget. 

\begin{table*}
\caption{Error budget (in nm rms). Comparison with simulations. Ratio $=IR Image SR / exp(-(2\pi \sigma_{ErrIR}/\lambda)^2)$}  
\label{err}
\centering   
\begin{tabular}{lcccccc}
	\hline
	\hline
 Asterism & \multicolumn{2}{c}{A47 } &  \multicolumn{2}{c}{A53} &   \multicolumn{2}{c}{A12} \\
 	\hline
  IR image hour &   \multicolumn{2}{c}{\hms{0}{10}{00}}  & \multicolumn{2}{c}{\hms{3}{26}{33}} &\multicolumn{2}{c}{\hms{6}{06}{41}} \\
  Engaged slopes hour &   \multicolumn{2}{c}{\hms{0}{10}{43}}  & \multicolumn{2}{c}{\hms{3}{24}{44}} &\multicolumn{2}{c}{\hms{6}{07}{07}} \\
           $r_0$ (cm)   &      \multicolumn{2}{c}{16.3 }    &    \multicolumn{2}{c}{10.0 }      & \multicolumn{2}{c}{13.0 }  \\
        \hline
              &          on-sky    &      simul  &     on-sky    &      simul       & on-sky        &      simul        \\
	$\sigma_{Tomo}$ &    156     &   161   & 219   &    220      &      188      &      201   \\
	 $\sigma_{OL}$  &        55    &     0   &    140     &    0         &          116      &    0           \\
	 $\sigma_{TomoNoiseFilt}$   &      48     &    78    &      56   &        79     &         97       &         156      \\
 	  $\sigma_{AliasGround} $&    95        &   95     &     132    &       132      &      83          &      83         \\
	 $\sigma_{AliasAlt}$    &       15     &    15   &      28  &      28       &        22        &          22     \\
	$\sigma_{BW}$   &      115      &    101   &     142     &     162        &        128        &        145       \\
	 $\sigma_{Fit} $  &     138       &     138   &     206    &    206           &         165       &       165         \\
	 $\sigma_{Field Stat} $  &    77    &     0   &   106      &       0        &        72        &          0        \\
         $\sigma_{NCPA}$&     115       &   0     &     115    &    0         &       115         &          0       \\
	\hline
	{ $\sigma_{ErrIR} $  } &     { 297 }     &    { 265}     &    {419}     &       {376}       &        {357}        &         {347}       \\
        \hline
	$\exp(-(2\pi \sigma_{ErrIR}/\lambda)^2)$ ($\%$) &     22.6       &    30.6    &    5.2     &        9.1    &         11.7       &        13.1       \\
	IR Image SR ($\%$) &       20.1     &     33.3   &      10.3   &       16.5      &       16.4     &          18.4     \\
	\hline
	{ Ratio} &      1.12     &     0.92   &      0.50   &       0.55      &        0.71      &          0.71     \\
\end{tabular}
\end{table*}

\subsection{Simulated error budget}
In this section we discuss the simulated error budget (columns \emph{simul} in Table~\ref{err}).
\begin{itemize}
\item 
We computed a 3-layer tomographic estimator, directly using the parameters of the simulation.
The tomographic error found while using it is $\sigma_{Tomo} =$  161, 220 and 201~nm rms. 
These values are in good agreement with the on-sky results 156, 219 and 188~nm rms. 
We recall that although the estimator is based on 3 layers, we introduced a 15 layers $C_n^2(h)$ profile in the simulated model of the atmosphere. Therefore in this particular case a representative tomographic model error was simulated. Table~\ref{TomoSimu} compares the performance of the 3-layer  to the 15-layer estimator within the numerical simulation. The performance when correcting  15 layers gives 128, 161 and 157~nm rms, a significant reduction of the error. As the reconstructed profile perfectly matches the simulated one, it represents the pure tomographic error $\sigma_{PureTomo}$ without any turbulence profile modelling error contribution ($\sigma_{Model}$).  The latter can be deduced by quadratically subtracting the tomographic performance of 15-layer estimator from the 3-layer one. This gives an estimation of the turbulence modeling error of $\sigma_{Model}$= 99, 150, 126~nm rms. These numbers can be directly compared with those in Table~\ref{TomoSky}, where we performed the same analysis with two estimators based on 3 or 15 layers, but with the on-sky data. The error $\sigma_{Model}$ is roughly a factor of 2 higher in the simulation which means we are only able with the simulation to estimate the order of magnitude of the model error that impact our observational results. In any case,  identifying only three layers to compute the estimator was clearly not sufficient and leads to around 120~nm of additional error. This significant contribution to the error budget underlines the requirement to increase the number of layers to be considered in the $C_n^2(h)$ profile. 

\begin{table}
\caption{Tomographic error measured from numerical simulations (in nm rms). The model error is deduced by subtracting the 3 layer performance to the 15 layer one.}  
\label{TomoSimu}
\centering   
\begin{tabular}{l c c c}
	\hline
	\hline
	      &       $\#$A47 & $\#$A53 & $\#$A12 \\
		\hline
		$\sigma_{Tomo}$ (3  layers)    &      161          &          220        &    201      \\
	$\sigma_{PureTomo}$ (15 layers)    &      128          &            161      &    157      \\
deduced $\sigma_{Model}$    &      99          &            150      &    126      \\
	\hline
\end{tabular}
\end{table}

\item 
The simulated noise injected in the loop $\sigma_{TomoNoiseFilt}$ is respectively 78, 79 and 156~nm rms. Notice here that we reproduced the same amount of noise per WFS in the slope measurements with respect to the one measured on-sky. Nevertheless, the noise propagated by the tomographic estimator is higher in simulations than on-sky. This difference is explained by the way the tomographic estimator was computed in the simulations. We did not include the slopes noise variance in the diagonal of the $C_{OffOff}$ covariance matrix to compute the tomographic estimator.
Therefore the simulated tomographic estimator fully propagates the noise, even though the noise is partially filtered by the integrator loop. 
\item 
For the bandwidth error, we find $\sigma_{BW}=$101, 162 and 145~nm rms with the considered conditions. The $\sigma_{BW}$ computed by simulation is close to the on-sky values.
\item 
As we use an analytical expression based on the value of $r_0$ to derive both  the aliasing and fitting error terms, we have reported the same number in the on-sky and simulations cases.
\item 
Open loop error, NCPA and field static aberrations errors were not simulated.
\end{itemize}
The total error expected on the IR image $\sigma_{ErrIR}$ from the simulated set of slopes is 265, 376 and 347~nm rms. The expected SR from the simulated slopes is therefore 30.6\%, 9.1\% and 13.1\% and must be compared to the SR measured on the simulated IR image which gives: 33.3\%, 16.5\% and 18.4\%. 
Here the global performance is better because some of the on-sky contributors were not considered in the simulation.

\subsection{Relative error budget}
\begin{figure}
\centering
\includegraphics[width=2.8cm]{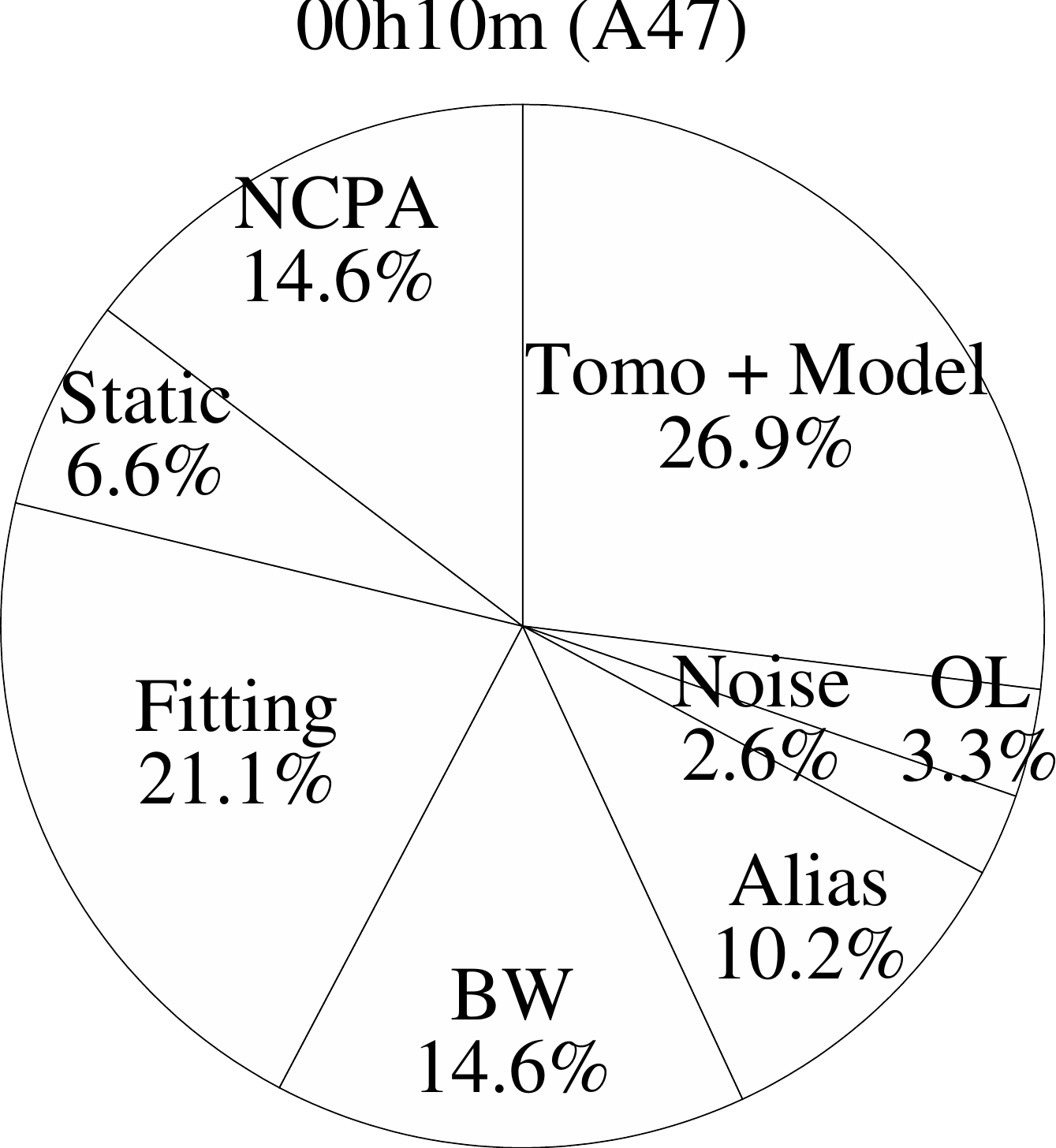}
\includegraphics[width=2.8cm]{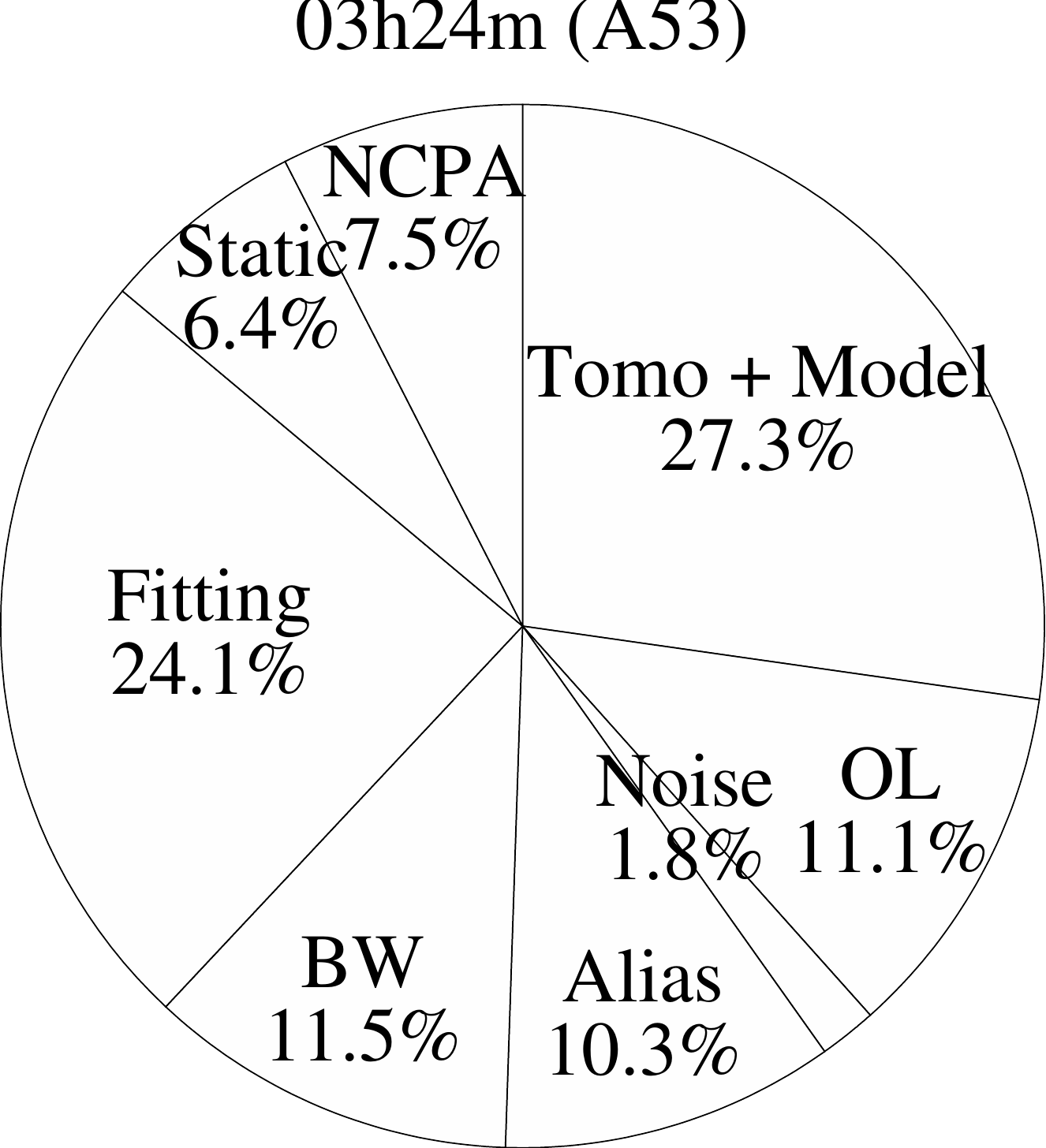}
\includegraphics[width=2.8cm]{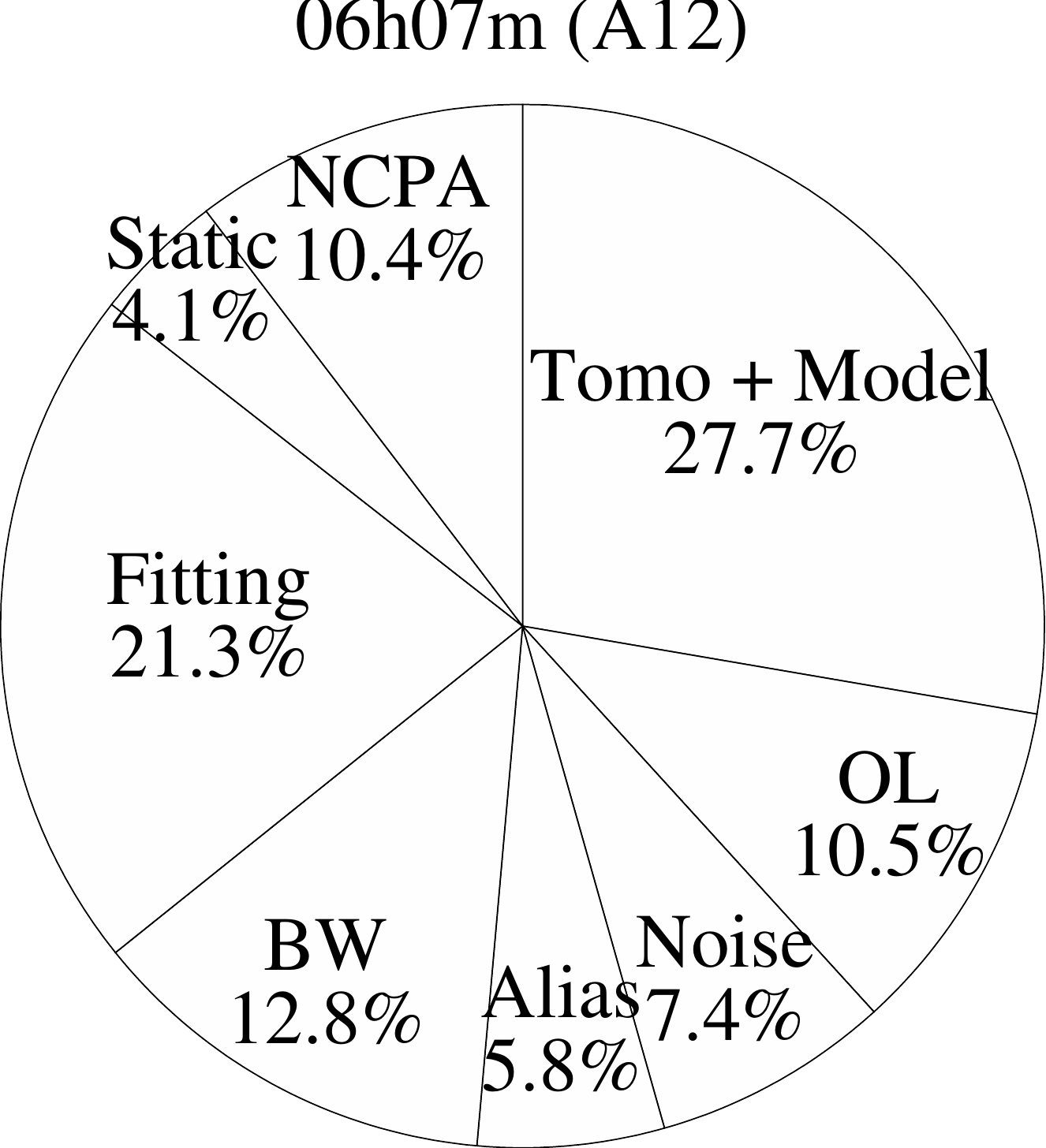}
\caption{Relative representation of the CANARY error budget at \hm{0}{10}, \hm{3}{24} and \hm{6}{07}. Areas of the chart are proportional to variances.}
\label{RelErrBudget}
\end{figure}

Fig.~\ref{RelErrBudget} shows the on-sky relative error budget (in \%) computed by the quadratic sum of the individual error terms presented in Table~\ref{err}.
\begin{itemize}
\item 
The fitting error counts for less than a quarter of the total error budget. This error is irreducible in CANARY as it is fixed by the number of actuators on the DM.
\item 
The tomographic error is the dominant term and its relative value remains remarkably stable at $\approx$ 27\% of the total error. This is the level of performance we expected from the initial simulations.  
Despite the change of guide star configuration and magnitudes, the estimator performs relatively well. $C_n^2(h)$ conditions were mainly dominated by a strong ground layer but MOAO provides better performance than GLAO. This demonstrates the robustness of our tomographic approach based on Learn and Apply. 
We emphasize again that because it is an on-sky measured term, it also includes the turbulence profile model error.
\item 
The next largest term in weight is the bandwidth error. It ranges between 11.5\% and 15$\%$. It is significantly larger than the noise contributor, showing that a higher sampling frequency could have been used. 
\item 
Noise injected in the loop contributes only a few percent of the total error ($\approx$ 2 to 7 $\%$) despite quite a large noise level in the individual WFS measurements (see Table~\ref{Simusparams}). 
\item 
The open loop term, also containing other unknown errors, ranges from $\approx$ 3 to 11 $\%$ of the total error. This term is at the level of the other contributors listed below. This a satisfactory demonstration that open loop operation works. 
\item
The aliasing relative contribution ranges between  6 and 10 $\%$ of the total error. 
\item 
The NCPA has a fixed value of 115~nm rms but its relative value evolves from 7.5$\%$ (strongest turbulence) to 14.7$\%$ (weakest turbulence).
\item 
The field static aberrations lead to a 4 to 6$\%$ relative error. This term was not expected at the beginning to have such a large contribution. It is clear that it will have to be properly mitigated in any MOAO system. 
\end{itemize}

\subsection{Discussion}

We would like to emphasize that the tomography terms measured on-sky match remarkably well with their corresponding numerical simulations. This is probably one of the most important results of the on-sky demonstration made by CANARY at phase A. Such a comparison makes us confident in establishing the error budget of future instruments and also in the ability to process the instrument data to properly retrieve  the $C_n^2(h)$ profile required for optimisation of the on-axis SR. 

The other main specific term of an MOAO instrument is the open loop error. From previous laboratory measurements a value of 3 to 5\% of the total turbulence perturbation was expected \citep{kellererDM2012}, leading here to 30 to 70~nm rms in our observing conditions. We measure on-sky values ranging from 50 to 160~nm rms. The difference may come from the way it has been evaluated. We measure it as the difference between the total error measured by the TS and the quadratic sum of all the error terms. 
Therefore the computed value can be influenced by any error in the estimation of the other terms. It contains the open loop error plus possibly all others unidentified errors such as  
drift in pupil alignment, badly seen turbulence on the bench etc...

Only one term has been identified on-sky that was not included in the initial simulations of CANARY ; the quasi-static field aberrations residuals. 
They are of the order of 70 to 130~nm rms and are very difficult to calibrate properly. For us, this term includes the error linked to the mis-calibrated field dependent aberrations and the creep of the DM linked to the static voltage offsets applied to it. On the E-ELT, those static field aberrations will also not be fully static, but evolve slightly because of the field derotation and the residual errors in the active optics. This will bring a problem to all tomographic techniques and thus will have to be solved for the E-ELT instruments. Particular effort on new calibration schemes should be envisioned to reduce this term and we emphasize here the importance of the Truth Sensor for such calibrations in CANARY.

The computed SR from the error budget shows an under-estimation of the measured SR on the IR camera. We explain the difference by several factors. Firstly, the $\exp(-(2\pi \sigma_{ErrIR}/\lambda)^2)$ formula tends to be pessimistic for SR below 30\%. We also observe this effect in the simulations. The simulations allow us to quantify the mismatch between the measured image SR and the exponential formula. The ratio of the IR image SR and the exponential formula is very similar for the simulations and the on-sky data as given in Table~\ref{err}. 
Secondly, we estimate the error bars of the total error budget to be $\pm  50$~nm rms leading to $\pm$4.0\% error on the predicted SR. Finally, we notice that $r_0$ fluctuated significantly (typically $\pm$2~cm in a few tens of seconds) between the recordings of the engaged slopes, the disengaged ones and the IR images. Unfortunately all the data were not perfectly synchronised. This may explain some discrepancies in Table~\ref{err}. 

Taking into account all these facts, we can say that the proposed error budget is validated.

\section{Conclusion}
\label{conclusion}
CANARY is a single channel MOAO demonstrator installed at the William Herschel Telescope. It obtained the first on-sky MOAO compensated images in September 2010. In this paper, we detail the calibration procedures used in CANARY, in particular to compute the optimized tomographic estimator required in MOAO and its software implementation. We present a method to process the slope data recorded on-sky and apply this to data taken during the night of the 27th of September 2010. The presence of the Truth Sensor allowed us to build a full error budget for the instrument. We evaluate 9 error terms  including the tomographic and the open loop error, which are new terms introduced by the MOAO control scheme. The tomographic error is estimated from synchronous disengaged slopes while the open loop error also requires data taken when the MOAO loop is engaged. We have also computed other classical AO error terms like noise, aliasing, bandwidth, fitting and NCPA. For the error budget, we distinguish two terms in the aliasing due to both the ground and the high altitude layer contributions. We check and successfully compare the measured on-sky error budget with numerical simulations for 3 datasets differing in guide star configuration, magnitudes and atmospheric conditions. The conclusion is encouraging, with the open-loop term, although larger than expected, being small compared to the other terms, and the tomographic term behaving as expected from simulation. But we identify on-sky an additional term linked to the residual quasi-static field aberrations found to be of the order of the open-loop term. 

The turbulence profile derived from the instrument data is presented and used in the tomographic optimization. The optimized {\em Learn \& Apply} tomographic estimator based on 3 layers computed during the on-sky operations performed at any time better than a basic ground layer reconstruction. The average performance of the 3 layer tomographic reconstruction was 216~nm rms while ground layer reconstruction performed at 270~nm rms representing an increase in the performance of 162~nm rms on average during the night. We are also able to evaluate the impact of the $C_n^2$ profile model error in the tomographic error by testing (in simulation) estimators computed with 3 or 15 layers. 
Although the tomographic error is the largest contributor to the error on CANARY, it remained stable contributing one quarter of the total error budget. The Strehl ratio on the IR camera ($\lambda=1\,530$~nm) was measured and compared to the estimated error budget. The agreement was good. In MOAO mode we measured SR in the range of 10\% to 20\%  with a typical r$_0$ value between 10 to 16~cm. Comparison with the numerical simulation makes us confident in the on-sky error budget estimation. Therefore this paper brings some important insights for the establishment of the wavefront error budget of future MOAO instruments. 

CANARY represents a significant advance in the implementation of  the future tomographic AO systems like MOAO, LTAO and even MCAO instruments. It has successfully demonstrated  the tomographic optimization in a direction of interest and the accuracy of the novel open loop control scheme. The next phase of CANARY uses 4 Rayleigh Laser Guide stars in addition of the 3 natural guide stars. This phase also implements a modification of the Learn \& Apply algorithm to perform with mixed measurements of natural and laser guide stars. 

\section*{Acknowledgments}
This work was supported by Agence Nationale de la Recherche (ANR) program 06-BLAN-0191, CNRS / INSU, Observatoire de Paris, and Universit\'e Paris Diderot Paris 7 in France, Science and Technology Facilities Council, University of Durham and the Isaac Newton Group of telescopes in UK and European Commission Framework Programme 7 (E-ELT Preparation Infrastructure Grant 211257 and OPTICON Research Infrastructures Grant 226604).

\section{Appendix}

\subsection{White noise propagation through a first-order low-pass filter}
\label{append_g/2-g}
The expression of the first-order low-pass filter is
\begin{equation}
\label{lowpass}
y_n = (1-g)y_{n-1} + g x_n\ .
\end{equation}
We want to determine the variance $\sigma_y^2$ of the output $y_n$, when $x_n$ is a white noise of known variance $\sigma_x^2$. We have
\begin{equation}
y_n^2 = (1-g)^2 y_{n-1}^2 + g^2 x_n^2 + 2g(1-g)y_{n-1}x_n\ .
\end{equation}
As $y_{n-1}$ and $x_n$ cannot be correlated, it follows that
\begin{equation}
\sigma_y^2 = (1-g)^2 \sigma_y^2 + g^2 \sigma_x^2
\end{equation}
or finally
\begin{equation}
\label{cqfdg2g}
\sigma_y^2   = \frac{g}{2-g} \sigma_x^2\ .
\end{equation}

\subsection{White noise propagation through a first-order low-pass filter with fractional shift}
\label{append_gaga}
We let $y_n$ be a signal resulting from the low-pass filtering described in Eq.~\ref{lowpass}, of a white noise $x_n$ of variance~$\sigma_x^2$.
We consider the signal $s_n$ deduced from $y_n$ by a fractional delay:
\begin{equation}
s_n = \alpha y_{n-2} + (1-\alpha) y_{n-1}\ .
\end{equation}
We can replace $y_{n-1}$ in the above equation with its expression in terms of $y_{n-2}$ using Eq.~\ref{lowpass}, and compute the square of $s_n$:
\begin{eqnarray}
s_n^2 & =   &  (\alpha y_{n-2} + (1-\alpha) y_{n-1})^2    \\ 
          & =  &  ( \alpha y_{n-2} + (1-\alpha) (1-g) y_{n-2} + (1-\alpha) g x_{n-1})^2    \\ 
          &   = &  ((1-g+\alpha g)y_{n-2} +  (1-\alpha) g x_{n-1})^2   \\
          &   = &  (1-g+\alpha g)^2  y_{n-2}^2 + (1-\alpha)^2 g^2 x_{n-1}^2 \\
          &      &  + x_{n-1} y_{n-2} (1-g+\alpha g) (1-\alpha) g \ .
\end{eqnarray}
Considering that $x_{n-1}$ and $y_{n-2}$ are not correlated, and using Eq.~\ref{cqfdg2g} for expressing the variance of $y_{n-2}$, it follows that
\begin{eqnarray}
\sigma_s^2 & = & \left( (1-g+\alpha g)^2 \frac{g}{2-g} + (1-\alpha)^2 g^2 \right) \sigma_x^2  \\
                  & = & \frac{g}{2-g} (1 - 2 g \alpha (1-\alpha)) \, \sigma_x^2\ .
                  \label{cqfdag}
\end{eqnarray}
The re-arrangement of terms as in Eq.~\ref{cqfdag} makes it clear that the result is symmetric between $\alpha$ and $(1-\alpha)$. Values of $\alpha$ of either 0 or 1 (integer shift) reduce the expression to the previous case, with a filtering coefficient of $g/(2-g)$. The particular case of $\alpha=0.5$ leads to $\sigma_s^2 = ( g/2) \, \sigma_x^2 $.

Now, the variance of the difference between $x_n$ and $s_n$ can easily be deduced from the previous calculus, because $x_n$ and $s_n$ cannot be correlated. The variance of their difference is just equal to
\begin{equation}
\sigma_{x-s}^2 = \left(1+\frac{g}{2-g} (1 - 2 g \alpha (1-\alpha)) \right) \, \sigma_x^2
\end{equation}

\bibliographystyle{aa} 
\bibliography{canary_PhaseA_AA_FINAL} 
\end{document}